\documentclass[sigconf]{acmart}
\PassOptionsToPackage{dvipsnames}{xcolor}

\PassOptionsToPackage{colorlinks=true,linkcolor=RubineRed,breaklinks=True,citecolor=RubineRed}{xcolor}
\usepackage{algorithm, algorithmicx, algpseudocode}
\usepackage{enumitem}
\setlist[itemize]{noitemsep, topsep=0pt}

\usepackage[most]{tcolorbox}
\usepackage{adjustbox}
\usepackage{multirow}
\usepackage{booktabs,subcaption,dcolumn}
\usepackage{tikz}
\usepackage{enumitem}
\usepackage{tabularx}
\usepackage{MnSymbol}

\usepackage{pifont}
\usepackage{svg}
\usepackage{soul}
\usepackage{amsmath}
\usepackage{dirtytalk}
\usepackage{subcaption}
\usepackage{graphicx}
\usepackage{setspace}
\usepackage{caption}
\usepackage{booktabs} 
\usepackage[titletoc,toc,title]{appendix}
\usepackage{svg}
\usepackage{amsfonts}
\usepackage{xurl} 

\usepackage{diagbox}
\usepackage{indentfirst}

\usepackage{CJKutf8}
\usepackage{amsmath}
\usepackage{amsfonts} 

\usepackage{colortbl}

\copyrightyear{2025}
\acmYear{2025}
\setcopyright{acmlicensed}
\acmConference[CODASPY '25]{Proceedings of the Fifteenth ACM Conference on Data and Application Security and Privacy}{June 4--6, 2025}{Pittsburgh, PA, USA}
\acmBooktitle{Proceedings of the Fifteenth ACM Conference on Data and Application Security and Privacy (CODASPY '25), June 4--6, 2025, Pittsburgh, PA, USA}
\acmDOI{10.1145/3714393.3726490}
\acmISBN{979-8-4007-1476-4/2025/06}

\AtBeginDocument{%
  \providecommand\BibTeX{{%
    \normalfont B\kern-0.5em{\scshape i\kern-0.25em b}\kern-0.8em\TeX}}}

\newcolumntype{?}{!{\vrule width 1.5pt}}

\newcommand{\textbox}[1]{
    \noindent\fbox{%
        \parbox{0.97\columnwidth}{%
            {#1}
        }%
    }
}

\newtcolorbox{cooltextbox}[1][]{%
    colback=black!5,
    colframe=black!5,
    notitle,
    sharp corners,
    borderline west={0pt}{0pt}{red!80!black},
    enhanced,
    breakable,
    left=0pt,
    right=0pt,
    top=0pt,
    bottom=0pt
    }

\newcommand\smamath[1]{{\small $#1$}}
\newcommand\smacal[1]{{\small $\mathcal{#1}$}}

\newcommand\ftmath[1]{{\footnotesize $#1$}}

\newcommand\scmath[1]{{\scriptsize $#1$}}

\newcommand\revision[1]{%
  \bgroup
  \hskip0pt\color{blue!80!black}%
  #1%
  \egroup
}

\newcommand\todo[1]{%
  \bgroup
  \hskip0pt\color{red!80!black}%
  #1%
  \egroup
}

\newcommand\fmw[1]{{\fontfamily{pcr}\selectfont {\small ATS-SF}}}
\newcommand\scfmw[1]{{\fontfamily{pcr}\selectfont {\scriptsize ATS-SF}}}
\newcommand\ftfmw[1]{{\fontfamily{pcr}\selectfont {\footnotesize ATS-SF}}}

\begin{document}
\title{The Ephemeral Threat: Assessing the Security of Algorithmic Trading Systems powered by Deep Learning}

\author{Advije Rizvani}
\email{advije.rizvani@uni.li}
\orcid{0009-0001-7189-5660}
\affiliation{%
  \department{Liechtenstein Business School}
  \institution{University of Liechtenstein}
  \city{Vaduz}
  \country{Liechtenstein}
}

\author{Giovanni Apruzzese}
\orcid{0000-0002-6890-9611}
\email{giovanni.apruzzese@uni.li}
\affiliation{%
  \department{Liechtenstein Business School}
  \institution{University of Liechtenstein}
  \city{Vaduz}
  \country{Liechtenstein}
}

\author{Pavel Laskov}
\email{pavel.laskov@uni.li}
\orcid{0000-0002-3212-7167}
\affiliation{%
  \department{Liechtenstein Business School}
  \institution{University of Liechtenstein}
  \city{Vaduz}
  \country{Liechtenstein}
}

\begin{abstract}
We study the security of stock price forecasting using Deep Learning (DL) in \textit{computational finance}. Despite abundant prior research on vulnerability of DL to adversarial perturbations, such work has hitherto hardly addressed practical adversarial threat models in the context of DL-powered \textit{algorithmic trading systems}~(ATS).

Specifically, we investigate the vulnerability of ATS to adversarial perturbations launched by a realistically constrained attacker. We first show that existing literature has paid limited attention to DL security in the financial domain---which is naturally attractive for adversaries. Then, we formalize the concept of \textit{ephemeral perturbations} (EP), which can be used to stage a novel type of attack tailored for DL-based ATS. Finally, we carry out an end-to-end evaluation of our EP against a profitable ATS. Our results reveal that the introduction of small changes to the input stock-prices not only (i)~induces the DL model to behave incorrectly but also (ii)~leads to the whole ATS to make suboptimal buy/sell decisions, resulting in a worse financial performance of the targeted ATS. 
\end{abstract}

%
%



\maketitle

\section{Introduction}
\label{sec:introduction}

\noindent
In recent years, Deep Learning (DL) has become a popular tool in the financial sector, transforming many aspects of data analysis and decision-making processes~\cite{huang2020deep,heaton2017deep,ozbayoglu2020deep}. One of the standout strengths of DL models is their ability to excel in time-series forecasting, which is essential for predicting market trends and guiding investment strategies~\cite{sezer2020financial}. In practice, these models are now integral to comprehensive systems that automate the entire ``predict--buy/sell--repeat'' process, known as Algorithmic Trading Systems (ATS)~\cite{swiss2021ai}. ATS have profoundly reshaped the financial landscape by determining the timing, type, and volume of trades~\cite{wang2021survey}. With the integration of DL, these systems (see Fig.~\ref{fig:ats}) execute transactions at scales and speeds beyond human capabilities, thereby minimizing emotional biases in trading decisions~\cite{hansen2020virtue}.

However, the reliance on DL models in such critical applications is not without risks. These models are known to be vulnerable to adversarial perturbations—small, carefully crafted changes to input data that can lead to incorrect predictions~\cite{biggio2018wild}. In the fast-paced and competitive world of financial trading, such errors can result in serious consequences. What if an attacker can stealthily introduce small and \textit{short-lived} perturbations to the data analysed by an ATS? Indeed, an attacker may tamper with a single data-point (the effects of which are unknown to the attacker---she cannot see the future!), but the perturbation will be ``overwritten'' shortly afterwards with the new (clean) data issued by the broker: a persistent influx of perturbed data would be spotted by the organization using the ATS. 
Such a threat model, denoting a realistically feasible attack, has never been investigated before in this domain.

Indeed, as we will show, there is a lack of works that provide an holistic perspective of the DL-specific risks to ATS. For instance, some papers studied attacks against ``time-series forecasting'' which is a problem that can be solved with DL and that is very relevant for ATS~\cite{gallagher2022investigating, nehemya2021taking}. However, the way such attacks are evaluated resembles the typical viewpoint of ``adversarial machine learning papers'', i.e., the attack is deemed successful depending on how well it ``fooled'' the \textit{targeted DL model}~\cite{nehemya2021taking}. We argue that such an evaluation procedure is not only simplistic in the general sense (it is well known that DL models are a mere component of a much larger system~\cite{apruzzese2023real}), but also that an appropriate assessment of an attack against ATS must stem from analysing \textit{the gains/losses of the system} that occur as a consequence to the DL-model's output. 

Hence, in this work we formalize the concept of \textbf{ephemeral perturbation}~(EP), which are particularly designed to attack the DL models embedded in ATS. Then, we empirically assess the effects of EP against a representative ATS, built through an original ``ATS Security Framework''. We use publicly available data~\cite{yahooFinance} to develop practical DL models for time-series forecasting, which are integrated into our ATS. After validating the performance of our ATS and ensuring it demonstrates a reasonable level of profitability under typical market conditions, we proceed to introduce various EP to the system. We assess the impact of these perturbations by considering their effects on both {\small \textit{(i)}}~the DL models in isolation and {\small \textit{(ii)}}~the overall ATS. Our findings reveal that EP can subtly degrade the profitability of the ATS. Intriguingly, the degradation can be so small that the ATS still leads to a  gain---but \textit{inferior} to the gain without any EP: hence, the targeted organization will persist in using the ATS, thereby (unknowingly) \textit{decreasing their returns} over~time.

\underline{\textbf{\textsc{C}{ONTRIBUTIONS.}}} This paper explores the security risks associated with DL-powered ATS, a critical yet underexplored area in both security and financial computing literature. Specifically, we:
\begin{itemize}[leftmargin=*,noitemsep,topsep=0pt]
    \item highlight the oversight of financial applications in research focused on security of artificial intelligence~(\S\ref{sec:related});
    \item introduce and formalize the concept of ephemeral perturbations~(EP), which can realistically affect DL-based ATS~(\S\ref{sec:threat});
    \item propose a framework for security assessment of ATS~(§\ref{sec:ats}) which we use to study the impact of EP on an exemplary ATS (\S\ref{sec:assessment}, \S\ref{sec:additional}).
\end{itemize}
We also discuss and validate our findings with an user study with experts (§\ref{sec:discussion}). To foster future research, we release our resources~\cite{ourRepo}.

\section{Related Work}
\label{sec:related}

\noindent
We first define the problem space of our paper by summarizing prior related work~(§\ref{ssec:background}), and then carry out a systematic literature review wherein we pinpoint the research gap tackled by this paper~(§\ref{ssec:gap}).

\subsection{Background}
\label{ssec:background}

\noindent
Our paper tackles two orthogonal research fields gravitating around artificial intelligence (AI): applications of AI in finance, and security of AI. We stress that the notion of ``AI'' encompasses both machine learning (ML) and deep learning (DL) methods.

\vspace{1mm}
\noindent
\textbf{AI in Finance.}
There is a deluge of ways to use AI in finance. An early review of various AI-based techniques is the 2010 paper by Bahrammirzaee~\cite{bahrammirzaee2010comparative}. Since then, however, this field has substantially advanced, as shown by more recent summarizations~\cite{hilpisch2020artificial,dixon2020machine,goodell2021artificial}. In particular, the advent of deep learning (DL) in this domain caused a paradigm shift~\cite{huang2020deep,ozbayoglu2020deep} due to its demonstrated superiority over traditional machine learning (ML) or statistical techniques (for, e.g., forecasting the stock market~\cite{siami2018comparison}). Noteworthy applications of AI in finance, for which there is evidence of real-world deployment~\cite{swiss2021ai}, include: algorithmic trading~\cite{gomber2018algorithmic,le2018deep}, portfolio management~\cite{hu2019deep}), credit scoring~\cite{gunnarsson2021deep}, fraud detection~\cite{mishra2018credit,palaiokrassas2024leveraging_fc}. In this paper, we focus on the former: algorithmic trading systems (ATS) are a flourishing technology in the current financial landscape. ATS can greatly benefit from AI (and, specifically, from DL) thanks to its capabilities of predicting future market values~\cite{wang2021survey,hansen2020virtue,koki2022exploring_fc}. Such forecasts can then be used by the ATS to ``quickly'' make informed decisions on whether to buy/hold/sell a given commodity. There is evidence suggesting that modern ATS do, in fact, rely on the forecasting capabilities of AI~\cite{quantiacs,starfetch}.

\vspace{1mm}
\noindent
\textbf{Security of AI.}
The never-ending advances of AI induced many researchers to investigate the security of learning-based algorithms. This research field, typically referred to as ``adversarial machine learning''~\cite{biggio2012poisoning,biggio2013evasion,szegedy2014intriguing,biggio2018wild}, literally exploded in the last decade, after the discovery that deep neural networks could be fooled by tiny perturbations in the input data~\cite{biggio2018wild}. Thousands of papers have hitherto studied how learning models can be affected by such ``adversarial perturbations'', which can target any given model either during its training or inference stage, and which envision threat models assuming attackers with various degrees of knowledge and capabilities~\cite{apruzzese2022wild}. Despite all such work, however, a pragmatic solution to ``adversarial attacks'' has yet to be found~\cite{apruzzese2023real}. Moreover, despite the numerous domains in which AI has found applications (in research or in practice), most prior literature on AI security considers attacks against models devoted to computer vision (e.g.,~\cite{zimmer2024closing_fc}), as highlighted in~\cite{apruzzese2023real}. Hence, it is difficult to gauge how much previously proposed attacks can ``practically'' affect models embedded in other types of systems. The only way to do so is by formalizing specific threat models and assessing their effects.

\subsection{Research Gap (Security of AI in Finance)}
\label{ssec:gap}

\noindent
Prior works have investigated various security aspects of computational finance (e.g., in decentralized settings or blockchain~\cite{guo2024pride_fc,ankile2023see_fc,truong2019towards_fc}), and some papers are related to our focus on automated trading (e.g.,~\cite{bartoletti2022maximizing_fc,zhou2021high_fc}). However, the security of AI-specific applications in finance has not been adequately scrutinised by prior work. This highlights a research gap that we aim to fill with our paper.\footnote{We stress that the field of ``market manipulation''~\cite{wang2020market} is orthogonal to our focus: our goal is to attack an ATS which leads to losses ``only'' to its owners; whereas market manipulation focuses on inducing changes that affect a wide-array of target groups.}

To provide evidence of such a gap, we first turn the attention to a recent work~\cite{apruzzese2023real} revealing that, in 2019--2021, only two papers on ``adversarial machine learning'' (published in top-tier security venues) considered financial data~\cite{nasr2021adversary,nasr2019comprehensive}. However, none of these works focused on the financial domain: both~\cite{nasr2021adversary,nasr2019comprehensive} used financial data as a yet-another benchmark to validate some well-known attacks, but do not make considerations on how much the corresponding attacks may affect the owners of the targeted model. Yet, to provide a compelling contribution emphasizing that the AI-specific security aspects in finance have not been well studied, we carry out a comprehensive analysis of prior works.

\subsubsection{\textbf{Systematic Literature Review (top-tier venues).}}
We begin by systematically reviewing prior works that have been published in top-tier security conferences (similarly to, e.g.,~\cite{arp2022and,apruzzese2023real}). First, we collected the 7,266 papers that appeared in 9 top-tier venues ({\small ACM CCS} and {\small AsiaCCS}, {\small IEEE S\&P} and {\small EuroS\&P}, {\small FC}, {\small ACSAC}, {\small ESORICS}, {\small NDSS}, {\small USENIX SEC}) held 2014--2023; we only considered full papers (and not, e.g., short or workshop papers). Then, we inspect the abstract of all these publications. The goal is identifying candidates that may be related to ``security of AI applications in finance''. We disentangle this goal by defining three criteria:
\begin{itemize}[leftmargin=*]
    \item \textit{AI-related}: we consider a paper to be (potentially) about ``AI applications'' if the abstract of the paper mentions at least one of the following terms: ``machine learning'', ``artificial intelligence'', ``deep learning'', ``AI'', ``ML'', ``DL''.
    \item \textit{Finance-related}: we consider a paper to be (potentially) about ``finance'' if the abstract of the paper mentions at least one of: ``finance'', ``econom'', ``trading''.
    \item \textit{Security-related}: we only considered security-related venues, so we are confident that the paper is about security. However, given that the security of AI is typically associated with the term ``adversarial'' (e.g., ``adversarial perturbations / examples / ML''), we consider a paper to be within our scope if the term ``adversarial'' is mentioned at least once in the abstract.
\end{itemize}
We develop a script to carry out this keyword search. 
Overall, we find only two papers that have matches for each criteria, namely:~\cite{jang2017objective,liu2018data}. Finally, we manually review each of these papers~\cite{jang2017objective,liu2018data} to ascertain whether it truly deals with the security of AI in finance (with a focus on ATS). This process was done by two authors who independently reviewed each paper and discussed their findings. We found that none of these works can be considered to be about ``security of AI in finance'', despite matching our criteria. Indeed, both~\cite{jang2017objective,liu2018data} are about generic AI security, and mention ``finance'' only as a potential application of AI and no finance-specific assessment is carried out. Notable examples of excluded papers (which do not meet all three criteria) are, e.g.,~\cite{zhang2020dstyle}, which deals with AI in finance (despite not mentioning any AI-related term in the abstract) and mentions ``adversarial'' to denote generative-adversarial networks, which is just a yet-another application of AI (i.e., it is not used to indicate a security assessment of AI); and~\cite{zhou2021high_fc}, which is about security in finance and mentions ``adversarial'', but is not about AI-specific security risks.

\subsubsection{\textbf{Extended Literature Review.}}
Such a ``negative result'' inspired us to carry out a larger investigation on Google Scholar, which we do by following three steps (which we perform once in Jan. 2024, and a second time in May 2024 for validation): 
\begin{enumerate}[leftmargin=*]
    \item \textit{Broad Search.} We search for \textit{peer-reviewed papers} (excluding, e.g., preprints) that match queries relevant for our scope. Specifically, we consider: ``{\small adversarial perturbations algorithmic trading}'', ``{\small deep learning stock forecasting adversarial attack}'', and four queries corresponding to ({\small ``adversarial attack''\smamath{\land}(``financial forecasting''\smamath{\lor}``algorithmic trading''\smamath{\lor}``high-frequency trading''\smamath{\lor} ``portfolio management'')}). Each of these six searches returned \scmath{>}10k results. 
    
    \item \textit{Preliminary Screening.} For each search query, we inspect the returned papers to verify if the search terms occur in the considered paper. 
    We inspected 390 papers in this way, i.e., until page 10 of the results returned by Google Scholar. (A preliminary check showed any paper beyond page 10 did not contain all the terms in the query.) Ultimately, we found that only 17 papers contained \emph{all terms} specified in the respective query. 

    \item \textit{Detailed Filtering.} We further analyse the 17 papers obtained from the previous step, to determine if they truly addressed ATS security. To this end, one author reviewed each paper and shared the corresponding conclusions with another author who acted as a validator. After this inspection, we identified only 6 papers which are relevant to security analysis of AI for ATS (e.g., we removed~\cite{koshiyama2020algorithms} because its contributions have no connection with security). We then applied the snowball method~\cite{wohlin2014guidelines} on these 6 papers, looking for works that cite (or are cited by) them but our results did not change. 
\end{enumerate} 
These six papers are:~\cite{chen2021adversarial,dang2020adversarial,mode2020adversarial,goldblum2021adversarial,nehemya2021taking,gallagher2022investigating}). 
Despite providing significant contributions, these six papers have two important limitations from a security viewpoint:
\begin{itemize}[leftmargin=*,noitemsep,topsep=0pt]
    \item \textit{Powerful Attacker}: 4 papers~\cite{chen2021adversarial,dang2020adversarial,goldblum2021adversarial,nehemya2021taking} envision an adversary with extraordinary knowledge (e.g., they know everything about the model) or capabilities (e.g., they can arbitrarily manipulate the input data): such assumptions may not reflect a realistic scenario~\cite{apruzzese2023real}, given that the internal components of ATS are kept secret, and indiscriminate manipulation of the input samples may be deemed anomalous by the ATS which would reject the input.
    \item \textit{Model only:} 5 papers~\cite{mode2020adversarial,dang2020adversarial,nehemya2021taking,goldblum2021adversarial,gallagher2022investigating} consider the effects of the attack on the model in isolation---without considering the impact on the whole system (which is a recommendation by~\cite{apruzzese2023real}). The only paper that carries out a system-wide evaluation is the work by Chen et al.~\cite{chen2021adversarial}, which considers reinforcement learning for portfolio management---which are orthogonal to ATS (our focus).
\end{itemize}  
Finally, we stress that only two papers~\cite{dang2020adversarial,nehemya2021taking} release their source-code: this remark motivates our first technical contribution~(§\ref{sec:ats}).

\begin{cooltextbox}
\textbf{Takeaway.} Despite the increasing usage of AI in ATS, the system-wide impact that adversarial perturbations can have on AI-based ATS is still unclear. Our contributions seek to provide a foundation for future work in this domain.
\end{cooltextbox}

\section{ATS Security Framework (\fmw{})}
\label{sec:ats}

\noindent
To the best of our knowledge (and as shown in §\ref{sec:related}), there is no publicly available and open-source implementation of an ATS that can be used to carry out security assessments. To fill this gap and promote future work in this domain, we propose our ``ATS Security Framework,'' \fmw{}, which we describe in this section.

\begin{figure}[!t]
\vspace{-1mm}
    \centering
    \includegraphics[width=0.99\columnwidth]{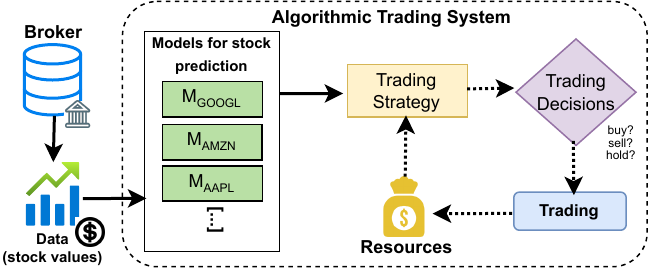}
    \vspace{-3mm}
    \caption{\textbf{Schema of an Algorithmic Trading System (ATS).}
    \textmd{\footnotesize The \textit{broker} (e.g., a bank) sends stock-related \textit{data} to a given organization which owns an ATS (dotted box). The ATS includes various DL models, used to make predictions on the basis of the input data. Such predictions are then used by the ATS to enact a given \textit{trading strategy}, which must account for the available \textit{resources} and decide what to do (i.e., buy/hold/sell). After making a decision, the resources are updated. }}
    \label{fig:ats}
    \vspace{-3mm}
\end{figure}

Our \fmw{} framework seeks to enable the development of exemplary ATS which can be used by researchers to simulate the workflow of a full-fledged ATS for security assessments.
We provide a schematic of our envisioned ATS in Fig.~\ref{fig:ats}, inspired by~\cite{jansen2018hands,quantiacs,backtrader}. 
The ATS (i.e., the rounded dotted box) receives data (i.e., the values of a certain set of stocks---e.g., {\small GOOGL} or {\small AMZN}) from a given \textit{broker} (e.g., a bank). Such data is then automatically analysed by various \textit{models}, which are collectively tasked to predict the ``future'' values of the stocks included in a given \textit{portfolio}. The output of such models is then used by the ATS according to a given \textit{trading strategy}, which will induce the ATS to make trading decisions (i.e., buy/hold/sell) on the stock market. Ideally, the decisions made by an ATS should yield a \textit{profit} to its owners---either in the short or long term. Such a profit can be measured either via {\small \textit{(a)}}~the ``cumulative daily returns'', which aggregates the gains/losses obtained after each day of use of the ATS~\cite{aloud2021intelligent,cumulativeReturn}; or via {\small \textit{(b)}}~the ``Sharpe Ratio''~\cite{riva2021learning,theate2021application}, which  measures the performance of an investment by adjusting for its risk, providing a ratio of return to volatility~\cite{sharpe1998sharpe}.

\vspace{1mm}
{\setstretch{0.8}
\textbox{{\small \textbf{Remark.} Our framework assumes that the ATS makes its decisions based on historical prices only. We are aware that there may be additional information sources (e.g., news and social media) that can be used to guide such trading decisions~\cite{gomez2019big,borch2022machine,colianni2015algorithmic,ferreira2021artificial}. Our implementation of \ftfmw{}, being open source, can be extended to account also for such data---but we leave development of such an enhancement to future work.}}}

\vspace{1mm}

\subsection{Low-level Architecture and Functionalities}
\label{ssec:ats_functionalities}

\begin{figure}[!htbp]
    \vspace{-3mm}
    \centering
    \includegraphics[width=0.99\columnwidth]{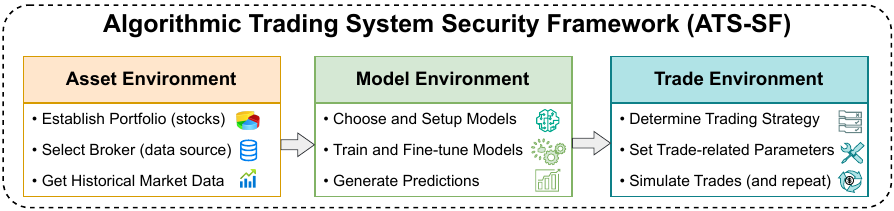}
    \vspace{-3mm}
        \caption{\textbf{Architecture of our \ftfmw{}.}
    \textmd{\footnotesize Our framework has three environments that allow fine-grained control on the entire management pipeline of an ATS, thereby enabling security assessments.}}
    \label{fig:fmw}
    \vspace{-3mm}
\end{figure}

\noindent
The design of our \fmw{} is centered around three core environments: Asset, Model, and Trade (summarized in Fig.~\ref{fig:fmw}). Such a design choice allows one to understand the effects of any change (potentially caused by an ``attack'') to each environment---enabling a system-wide assessment.
Let us explain how we devise our proposed \fmw{}, for which we must introduce some notation.

\noindent
\textbf{Asset Environment.}
First, the ATS assumes that its owners have specified a given portfolio, which we denote with \smacal{P}. Such a portfolio is identified by a set of stocks; let \smamath{s} denote each stock in a portfolio, so that \smamath{\bigcup_s = \mathcal{P}}. There exist various ways that can be used to select/optimize the stocks included in any portfolio (e.g.,~\cite{hu2019deep,tang2022asset_fc}). Obviously, the ATS expects to receive, as input, data (i.e., market values) pertaining to the stocks in \smacal{P} (obtainable from either public sources, such as Yahoo Finance~\cite{yahooFinance}; or from private brokers). Such data can come in many forms in terms of \textit{features} (e.g., opening or closing price of a given stock) and \textit{frequency} (e.g., the values can arrive on a daily basis, or even more often---the latter could be used for high-frequency trading\cite{sazu2022machine}). Our proposed \fmw{} enables the developer to freely choose any of these options.

\vspace{1mm}
\noindent
\textbf{Model Environment.}
Then, the ATS must analyse the input data and predict future values. To enable a wide-range of flexibility, we have realised \fmw{} so that it is \textit{model agnostic}: the predictions can be made, e.g., via deep learning methods (such as Long-Short Term Memory neural networks, or LSTM), or via traditional time-series forecasting techniques (e.g., Autoregressive Integrated Moving Average, or ARIMA), or even via other types of learning algorithms (tailored for regression). The developer has complete choice on how to implement such models: they can, e.g., perform multivariate analyses for each stock---such as using the opening and closing price to predict the closing price; or they can have a single model that takes as input the values of all the stocks in the portfolio, and predict a value for each stock. Of course, depending on the adopted choices, more operations may be required---such as \textit{training} the model(s) to fine-tune their parameters and optimize their performance. In what follows, we will use \smamath{M_s} to denote a model that is tasked to predict the values of stock(s) \smamath{s}.

\vspace{1mm}
\noindent
\textbf{Trade Environment.}
Finally, the ATS must make its trading decisions according to a given trading strategy (e.g.,~\cite{nuti2011algorithmic}), which must account for the output of the models \smamath{M_s} for all stocks \smamath{s\in\mathcal{P}}, the available resources (the ATS cannot ``buy'' any stock if there is no capital left), and any other contextual information (e.g., the historical trend of the stocks). The trading decisions are determined by providing rules that dictate the ``buy/sell/hold'' signals (i.e., if a buy signal is generated for a given stock \smamath{s} and there is sufficient available capital, the ATS will buy \smamath{s}). We also enable the developer to specify the portion of the portfolio used for the transaction (helpful for risk management). When making any given trade, the ATS accounts for the transaction fee~\cite{milionis2024automated_fc} as well as the slippage cost---both of which are user-specified. After the trade is made, our \fmw{} updates the available resources, accounting for gains and losses from the previous transactions as well as from the new values of the stocks in the portfolio. 

The entire procedure (i.e., receiving the input data, making the predictions, determining the trading decisions, executing the trades, updating the resources) is automatically repeated by our \fmw{} unless specified otherwise. At the end of the simulation, our \fmw{} will display the overall results of the ATS by showing: the predictions of each model, including how much they differ from the actual values---measurable, e.g., via the root mean squared error (RMSE); and the daily returns and Sharpe Ratio of the ATS.

\vspace{1mm}
{\setstretch{0.8}
\textbox{{\small \textbf{Disclaimer.} \ftfmw{} is meant to \textit{simulate} trading decisions over a pre-defined timespan, and does not guarantee that any parameter configuration will yield a net profit from the corresponding ATS (it is up to the developer to find the optimal configuration). Moreover, we do not claim that relying on our \ftfmw{} can lead to one gaining money from replicating the exact trading decisions---even if the results of the simulation show that the ATS consistently exhibits a remunerative Sharpe Ratio or daily returns. Hence, we \textit{do not} endorse using \ftfmw{} for real trading!}}}

\vspace{1mm}

\subsection{Custom ATS: Implementation \& Evaluation}
\label{ssec:ats_custom}

\noindent
We used our proposed \fmw{} to develop our custom ATS, which we will use as a baseline for our security assessment. We first describe how we developed the models and then explain how we developed the overarching ATS. Finally, we report the baseline performance of our prototype.

\vspace{1mm}
\noindent
\textbf{AI-models development.}
First, we collect data from Yahoo Finance~\cite{yahooFinance}, and we consider a timespan between Jan. 2010 and Oct. 2023. Then, we consider a portfolio \smacal{P} of 38 stocks, which is an holistic representation of the stock market. For each stock \smamath{s\in\mathcal{P}}, we develop a specific model, \smamath{M_s}; hence, our ATS encompasses 38 models. Each model relies on LSTM (due to their recognized capability for similar tasks~\cite{fischer2018deep}). To develop each model, we perform a temporal cut to our data by applying an 80:20 split (a common practice~\cite{kumar2021analysis}), identifying the training:testing partitions; moreover, each LSTM is tasked to predict the next ``close price'', done by means of a multivariate time-series analysis of 5 features (``high'', ``low'', ``open'', ``close'', ``volume'' -- all used in, e.g.,~\cite{ding2020study}) analysed over a temporal window of \smamath{w=50} days (used, e.g., in~\cite{mehtab2021stock}); we assume trades done on a daily basis. We then train all our models on the training set, trying out various configurations to optimize their performance (e.g., changing number of layers, or neurons per layer) and measure their performance on the test set: the RMSE (over the entire test set) is \smamath{4.03} on average (slightly better than that achieved by~\cite{kumar2021analysis}), confirming the quality of our models. 
We report the stocks in our portfolio and the RMSE achieved by each LSTM in Table~\ref{tab:stocks}.

\begin{table}[!htbp]
\centering
\vspace{-1mm}
\caption{\textbf{Stocks considered in the portfolio of our custom ATS.}
    \textmd{\footnotesize For each stock, we report the RMSE achieved by the its model over the test period. (Avg RMSE=4.03)}}
\label{tab:stocks}
\vspace{-3mm}
\resizebox{0.99\columnwidth}{!}{

\begin{tabular}{cc|cc|cc|cc|cc|cc}
\toprule

Stock & RMSE & Stock & RMSE & Stock & RMSE & Stock & RMSE & Stock & RMSE & Stock & RMSE\\
\midrule

{\small GOOGL} & \ftmath{6.36} & {\small AMZN} & \ftmath{4.52} & {\small AAPL} & \ftmath{4.55} & {\small BP} & \ftmath{1.17} & {\small BA} & \ftmath{6.83} & {\small MMM} & \ftmath{3.23} \\
{\small PEP} & \ftmath{3.48} & {\small JNJ} & \ftmath{2.50} & {\small PFE} & \ftmath{1.14} & {\small HON} & \ftmath{10.20} & {\small GE} & \ftmath{1.87} & {\small T} & \ftmath{0.44} \\
{\small MRK} & \ftmath{3.34} & {\small ABBV} & \ftmath{5.14} & {\small PG} & \ftmath{3.60} & {\small VZ} & \ftmath{0.85} & {\small TMUS} & \ftmath{2.89} & {\small HSY} & \ftmath{9.62} \\
{\small KO} & \ftmath{0.97} & {\small WMT} & \ftmath{1.47} & {\small JPM} & \ftmath{3.32} & {\small DUK} & \ftmath{2.27} & {\small SO} & \ftmath{1.98} & {\small EXC} & \ftmath{1.17} \\
{\small BAC} & \ftmath{1.11} & {\small GS} & \ftmath{14.11} & {\small V} & \ftmath{9.85} & {\small AEP} & \ftmath{1.70} & {\small AMT} & \ftmath{9.27} & {\small PLD} & \ftmath{3.49} \\
{\small XOM} & \ftmath{2.95} & {\small CVX} & \ftmath{4.34} & {\small COP} & \ftmath{3.37} & {\small SPG} & \ftmath{3.56} & {\small BHP} & \ftmath{2.19} & {\small RIO} & \ftmath{2.26} \\
{\small VALE} & \ftmath{0.69} & {\small FCX} & \ftmath{1.75} \\ 

\bottomrule
\end{tabular}
}
\vspace{-3mm}
\end{table}

\vspace{1mm}
\noindent
\textbf{Connecting the models to the ATS.}
After we have developed each model, we combine them together inside the ATS. To this end, we consider the well-known ``moving average crossover'' trading strategy~\cite{aycel2022new} which compares the predicted value (of the following day) with the previous predictions of each stock \smamath{s\in\mathcal{P}} to generate the trading decisions. Specifically, we use the previous predictions to calculate the moving average in the short-term (5 days) and in the long-term (20 days): a buy (sell) signal is given when the short-term average crosses above (below) the long-term average; otherwise, no signal is generated, defaulting to a hold decision. Depending on the signal, a trade can be made if there are enough resources: for our simulation, we set an initial capital of \smamath{100,000}\$ (common~\cite{sazu2022machine,riva2021learning}), and we specify a transaction cost of \smamath{0.005}\$ per share~\cite{IteractiveBroker}; the slippage cost is set at \smamath{0.02} to account for the difference between the expected and actual execution prices~\cite{lv2019empirical}.  Furthermore, to help managing risk and ensuring diversification whenever a trade is made, only a fixed portion of \smamath{10\%} of the current portfolio value is used. If no resources are available (i.e., if there is not enough capital for a ``buy'' signal, or there are no stocks for a ``sell'' signal), a signal is ignored. After each trading day, the portfolio is evaluated to calculate the daily returns, reflecting the changes (in percentage) of the portfolio value w.r.t. the previous day. Similarly, the Sharpe Ratio is calculated by comparing the excess return of the portfolio with its volatility, thereby evaluating the risk-adjusted performance of the portfolio; to this end, we set the risk-free rate to 5.05\% (given that our simulation spans over \smamath{\approx}2 years, and the 2-year treasury rate was 5.05\% according to~\cite{riskrate} on Oct. 23rd, 2023). Finally, to prevent lookahead bias, the trading signals are shifted backward by one day, so that they are based only on data available up to that day~\cite{baron2023measuring,wang2022information}.

\begin{figure*}[!t]
\vspace{-1mm}
\centering
  \begin{subfigure}[b]{0.32\textwidth}
    \centering
    \includegraphics[width=0.99\linewidth]{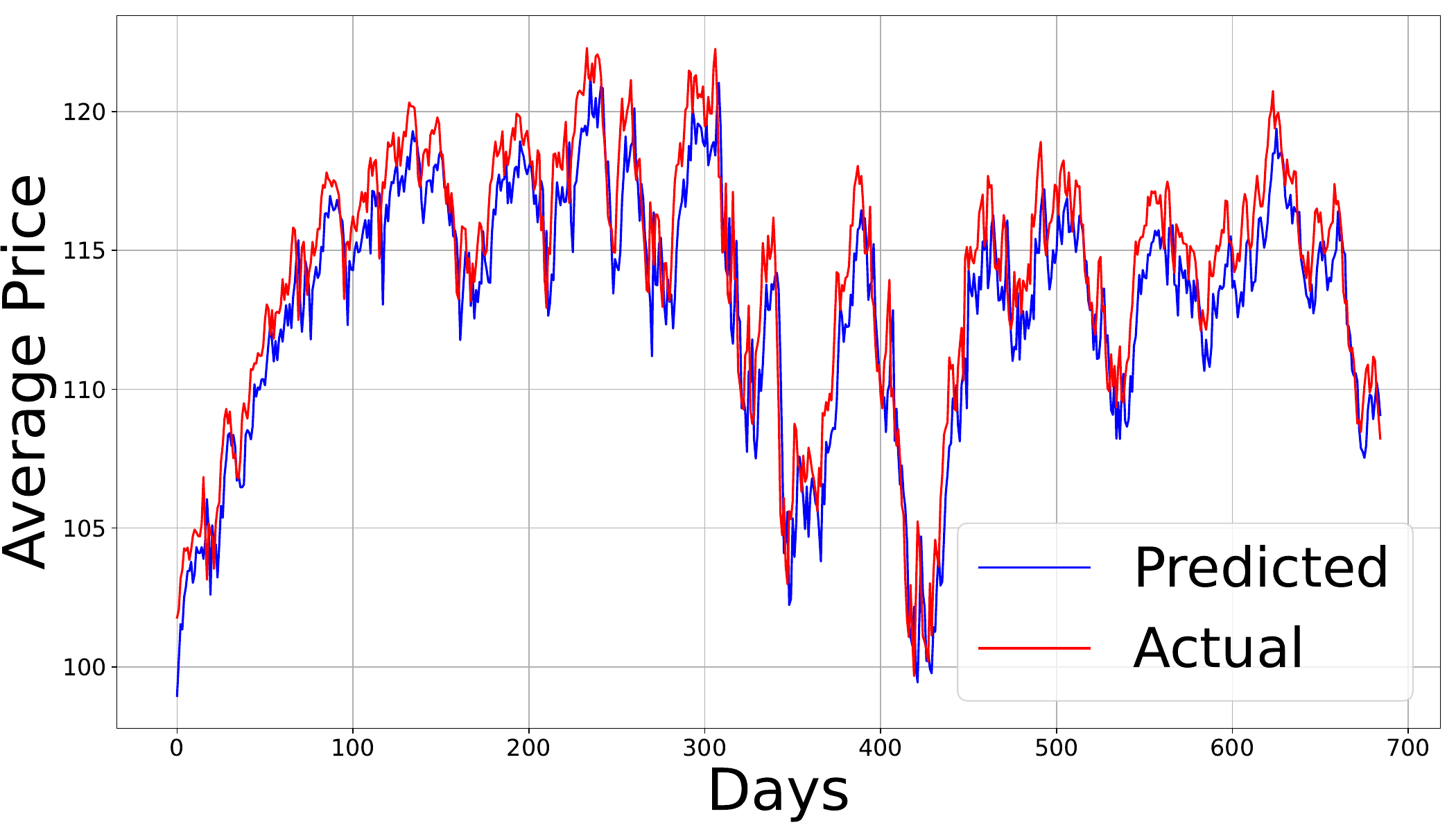}
    \caption{{\scriptsize \textbf{DL models.} (LSTM)}}
    \label{sfig:dl_baseline}
  \end{subfigure}
  \begin{subfigure}[b]{0.32\textwidth}
    \centering
    \includegraphics[width=0.99\linewidth]{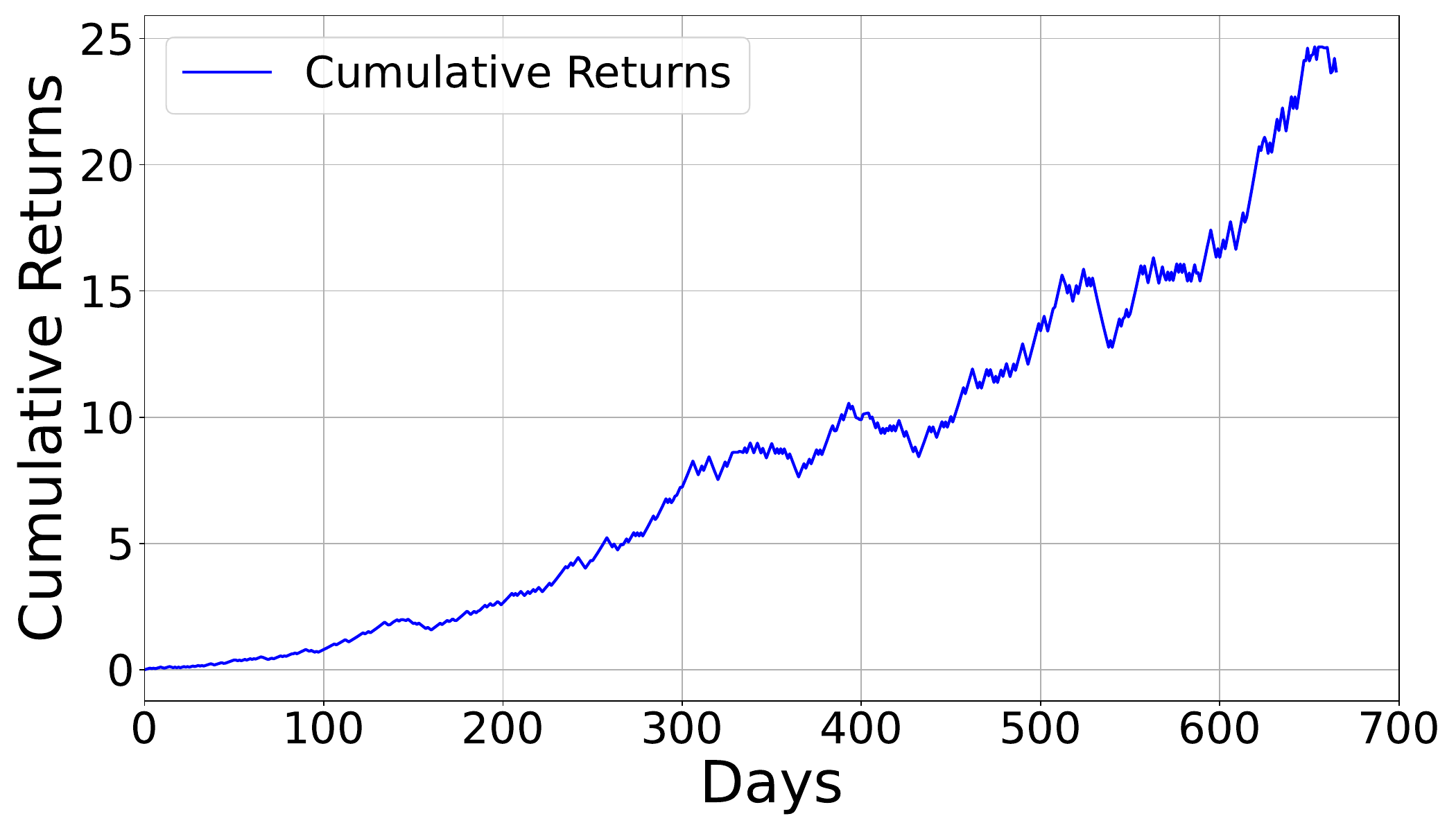}
    \caption{{\scriptsize \textbf{Cumulative Daily Returns.}}}
    \label{sfig:daily_returns_cumulative}
  \end{subfigure}
  \begin{subfigure}[b]{0.32\textwidth}
    \centering
    \includegraphics[width=0.99\linewidth]{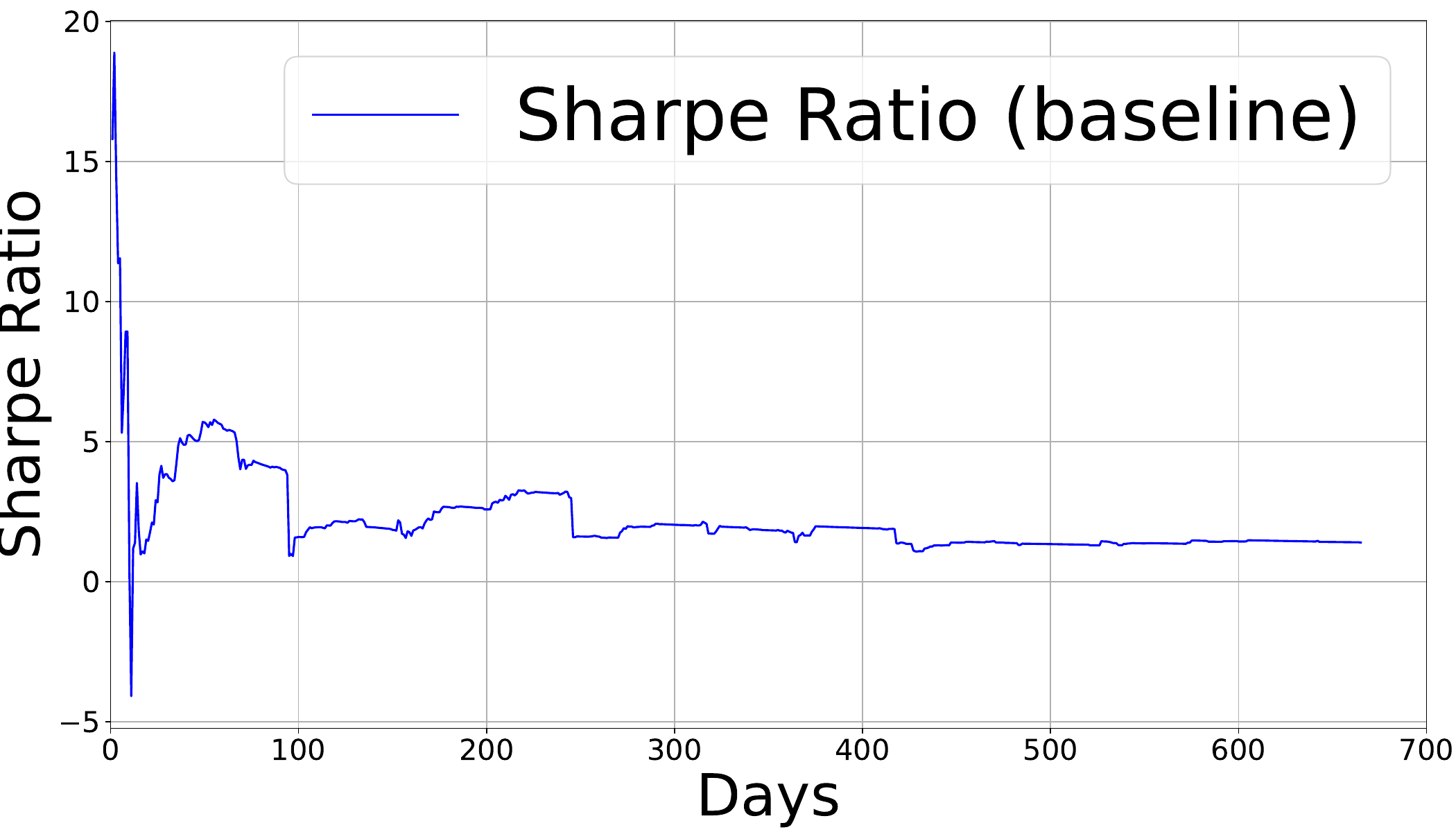}
    \caption{{\scriptsize \textbf{Sharpe Ratio.}}}
    \label{sfig:sharpe_ratio}
  \end{subfigure}
\vspace{-3mm}
\caption{\textbf{Baseline ATS.} \textmd{\footnotesize We show the profitability of our self-developed ATS. The LSTM models effectively predict (avg RMSE=3.89) the future close price of the stocks in the portfolio (Fig.~\ref{sfig:dl_baseline}). The ATS uses these predictions for its trading strategy, leading to trades that net a profit over-time, shown by increasing cumulative daily returns~(Fig.~\ref{sfig:daily_returns_cumulative}) and underscored by the Sharpe Ratio consistently above 0~(Fig.~\ref{sfig:sharpe_ratio}).}}
\label{fig:baseline}
\vspace{-4mm}
\end{figure*}

\vspace{1mm}
\noindent
\textbf{Performance Evaluation.}
We let our ATS simulate trades for the entire test window: from Dec. 2021 to Oct. 2023 (recall that we partition our 13 years of historical market data in train:test with a 80:20 split).
We report the baseline performance of our ATS in Fig.~\ref{fig:baseline}. Specifically, Fig.~\ref{sfig:dl_baseline} shows the aggregated performance of the LSTM models (for all 38 stocks in the portfolio), denoting that, overall, our models are accurate at predicting the closing price. Then, Fig.~\ref{sfig:daily_returns_cumulative} shows the cumulative daily returns over the entire ATS operation time, from which we see that the ATS is generating a profit of approximately 25\% over two years. Finally, Fig.~\ref{sfig:sharpe_ratio} shows the Sharpe Ratio: after the first days (which denote some stark fluctuations due to quick distribution of resources in the capital), the Sharpe Ratio of the ATS stabilizes and consistently remains above 0, denoting that the decisions made by the ATS are, ultimately, benefiting its potential owners---at~least according~to~our~simulation.

\vspace{-1mm}
\begin{cooltextbox}
\textbf{Takeaway.} Our ATS has an appreciable performance, which would justify its deployment in the real world (further validated by our user study in §\ref{ssec:expert}). Such a property makes our ATS a good test subject for our security assessment.
\end{cooltextbox} 

\section{Proposed Threat Model}
\label{sec:threat}

\noindent
Recall (§\ref{ssec:gap}) that prior related work envisioned threat models assuming powerful adversaries. As a result, the corresponding attacks were often very effective---at least from a ``model only'' perspective (e.g., the FGSM attack described in \cite{gallagher2022investigating} degraded the model’s performance by reducing its accuracy from 95\% to 60\%).

While such scenarios are not impossible, we argue that ``some'' real attackers may rely on easier and more subtle strategies to exploit the AI models deployed in ATS. We explain our vision by proposing our original threat model---and then compare it to the ``adversarial attacks'' envisioned in prior works (§\ref{ssec:compare}).

\subsection{Target System}
\label{ssec:target}

\noindent
The system (i.e., the ``defender'') targeted by the attacker resembles the exemplary ATS described in §\ref{sec:ats}. For completeness, we formalize its key points here.

We consider an ATS whose decisions are driven by data received by a given broker---which is considered to be trusted. Such data is in the form of historical market-related values (e.g., closing price, opening price), which pertain to a portfolio (\smacal{P}) encompassing a certain set of stocks; without loss of generality, we assume a daily frequency of broker-ATS communications~\cite{chen2021adversarial}. The ATS integrates models tasked to predict the future values of a specific stock \smamath{s\in\mathcal{P}}, so that \smamath{M_s} is the model devoted to predicting the values of \smamath{s}; without loss of generality, we assume that each stock has a specific model. 

A model \smamath{M_s} produces an output by analysing the previous \smamath{w} values of stock \smamath{s} in a time-series fashion: given a point-in-time \smamath{t}, \smamath{M_s} returns the prediction of \smamath{t+1} as: \smamath{M_{s}(t+1)=f(s,t,w)}, where \smamath{f} is the function learned by \smamath{M_s} during its training phase; ideally, \smamath{M_s(t+1)} should approximate the actual following value of \smamath{s}, i.e.,  \smamath{M_{s}(t+1) \approx s_{t+1}} (measurable via, e.g., RMSE~\cite{vijh2020stock}). The ATS takes all predictions of each model into account and then, depending on a given trading strategy and available resources, will enact certain trading decisions (e.g., buy/sell/hold any given stock). It is implicitly assumed that the ATS is expected to yield a profit to its owners (otherwise, they would not use it in the first place!).

\subsection{Envisioned Attacker}
\label{ssec:attacker}

\noindent
Our attacker has limited knowledge and capabilities, which implicitly increase the real-world likelihood of our threat model~\cite{apruzzese2023real}. 
\begin{itemize}[leftmargin=*,noitemsep,topsep=0pt]
    \item \textit{Knowledge.} The attacker know that the targeted ATS analyzes the historical market-data sent by the broker. However, the attacker does not know the entire portfolio considered by the ATS. Specifically, the attacker only knows a single stock \smamath{s'\in\mathcal{P}} (for instance, assuming the ATS considered in our implementation which has a portfolio with the 38 stocks in Table~\ref{tab:stocks}, the attacker will only know, e.g., that \smacal{P} includes {\small GOOGL}). Moreover, the attacker lacks any knowledge on the models (including \smamath{M_{s'}}) integrated in the ATS, including the length of the analysis period \smamath{w}.
    
    \item \textit{Capabilities.} The attacker can introduce some perturbations by manipulating the communications between the broker and the ATS (this is doable, e.g., via a man-in-the-middle attack---which are feasible in this context~\cite{nehemya2021taking}). However, to avoid being detected, such perturbations must be ``small'' (if the price of one stock is substantially different, the ATS may reject the input~\cite{ahmed2017anomaly}) and also ``short-lived,'' i.e., the attacker cannot send perturbations over many days (a single ``small'' deviation can be acceptable, but the owners of the ATS would react if they constantly receive wrong data from the broker). The attacker has no access to the ATS (which could be used for, e.g., query-based attacks~\cite{apruzzese2023real}). 
\end{itemize}
Our attacker has one goal: induce the targeted organization to \textit{gain less money}. This is a realistic assumption: given the competitive nature of the financial market, an adversary may want to gain an advantage by reducing the earnings of their competitors (even if such competitors keep gaining money). From the viewpoint of the ATS, such an effect can be measurable, e.g., by a Sharpe Ratio that, despite being inferior to the baseline value, is still above 0.

\subsection{Strategy: Ephemeral Perturbations}
\label{ssec:strategy}

\noindent
To reach their goal, our attacker has only one option: ``guess'' an \textit{ephemeral adversarial perturbation} (EP). Such an EP must be small enough to avoid raising suspicion---in the short-term (to avoid immediate reactions by the target organization) and in the long-term (if the ATS profitability drops excessively, then the organization would replace/stop using it). We formalize this strategy.

The attacker manipulates the values of the stocks that they know are analysed by the ATS, \smamath{s' \in \mathcal{P}}, which will likely lead to the corresponding model (\smamath{M_{s'}}) to output a different result, which may (or may not) induce the ATS to make a suboptimal trading decision. Our EP should last only a single time-point (i.e., only one day). Therefore, using an EP requires the attacker to make two choices:
\begin{itemize}[leftmargin=*,noitemsep,topsep=0pt]
    \item ``when'' the EP should be used---i.e., which day \smamath{t} to attack,
    \item ``how large'' the EP should be---i.e., what is the ``adversarial'' value\footnote{An attacker can modify \ftmath{s_t} in any way, but `large' changes (i.e., those leading to an \ftmath{s'_t} s.t. \ftmath{|s_t - s'_t| \gg 0}) would raise suspicion.} received by the ATS (and which will be analysed by \smamath{M_{s'}}) for the perturbed stock \smamath{s'\in\mathcal{P}}.
\end{itemize}  Hence, we can formally express an EP as a function: 
\begin{align}
    \text{EP}(s,t,m) = s'_t \neq s_t
    \label{eq:ep}
\end{align}
where \smamath{s_t} is the value of \smamath{s} at time \smamath{t}, \smamath{s'_t} is the adversarial value of \smamath{s_t} after the application of the EP, and \smamath{m} is a parameter which regulates the \textit{magnitude} of the perturbation.

\subsection{Comparison with prior threat models}
\label{ssec:compare}

\noindent
We have already established that, at a high-level, prior works that considered ``AI-specific attacks'' in the context of computational finance have limitations (§\ref{ssec:gap}). Here, we perform a low-level comparison of our proposed threat model with respect to those envisioned in the ten works we found during our literature review, namely:~\cite{nasr2021adversary,nasr2019comprehensive,jang2017objective,liu2018data,chen2021adversarial,dang2020adversarial,mode2020adversarial,goldblum2021adversarial,nehemya2021taking,gallagher2022investigating}.

First, we observe that our attacker shares some traits with the ``myopic'' attacker proposed in~\cite{apruzzese2022wild}, for which the targeted ATS is an ``invisible'' ML system~\cite{apruzzese2023real}. This is a crucial observation: our attacker \textit{cannot query the targeted model}, and \textit{does not know anything about the targeted model}. Such a consideration automatically puts our attacker in a different league than ``white-box'' attackers that rely on knowledge of the learned model gradients to craft an adversarial example (this is done, e.g., in~\cite{chen2021adversarial, nehemya2021taking, gallagher2022investigating, goldblum2021adversarial, jang2017objective, mode2020adversarial, nasr2019comprehensive, nasr2021adversary}). Moreover, even ``black-box'' attacks (envisioned in~\cite{nasr2021adversary}) which rely on querying the target model so as to craft a surrogate model that can then be used to apply the strategies for white-box attacks (by leveraging the transferability of adversarial examples~\cite{demontis2019adversarial}) cannot be staged by our attacker. We stress that Nehemeya et al.~\cite{nehemya2021taking} as well as Goldblum et al.~\cite{goldblum2021adversarial} also consider a ``black-box'' setting for their attack with no querying involved: they simply assess how much a perturbation to a given model can transfer to another model that uses a different algorithm/parameters. It is not explained (in either~\cite{nehemya2021taking,goldblum2021adversarial}) how an attacker can, in the real-world, obtain such a surrogate model; however, we have reason to believe that it requires substantial resources (e.g., an insider of a company using the ATS can obtain some information about the models' specifics).

Second, we observe that our attacker can only attempt a \textit{short lived} (and small) perturbation. This is substantially different from, e.g., the attack proposed by Dang et al.~\cite{dang2020adversarial}, in which it is stated that ``we suppose that the attacker is allowed to perturb all inputs at test time''. For instance, consider the models we developed for our custom ATS (§\ref{ssec:ats_custom}): each model receives as input a time-window of \smamath{w=50} values. Our attacker can only change one value (i.e., the last one) whereas the attacker envisioned by Dang et al.~\cite{dang2020adversarial} would be able to perturb all 50 values received as input by our models.

Third, our attacker can only operate during the \textit{inference phase} of the model. In other words, our attacker is not capable of staging ``poisoning attacks'' which require the manipulation of data used to train the model---as done by Liu et al., and also by Nasr et al.~\cite{liu2018data, nasr2021adversary}.

\vspace{1mm}
{\setstretch{0.8}
\textbox{{ \textbf{Simplicity.} Compared to previously proposed attacks, ours is much more simple to realize: our attacker \textit{knows nothing} of the targeted system, and only manipulates \textit{one value} sent by the broker. We are not aware of any previously proposed attack that leverages a similar concept as our ``ephemeral perturbations'' in the financial context---motivating our upcoming evaluation.}}
}

\begin{figure*}[!t]
\centering
    \begin{subfigure}[!htbp]{0.99\columnwidth}
        \centering
        \frame{\includegraphics[width=0.99\columnwidth]{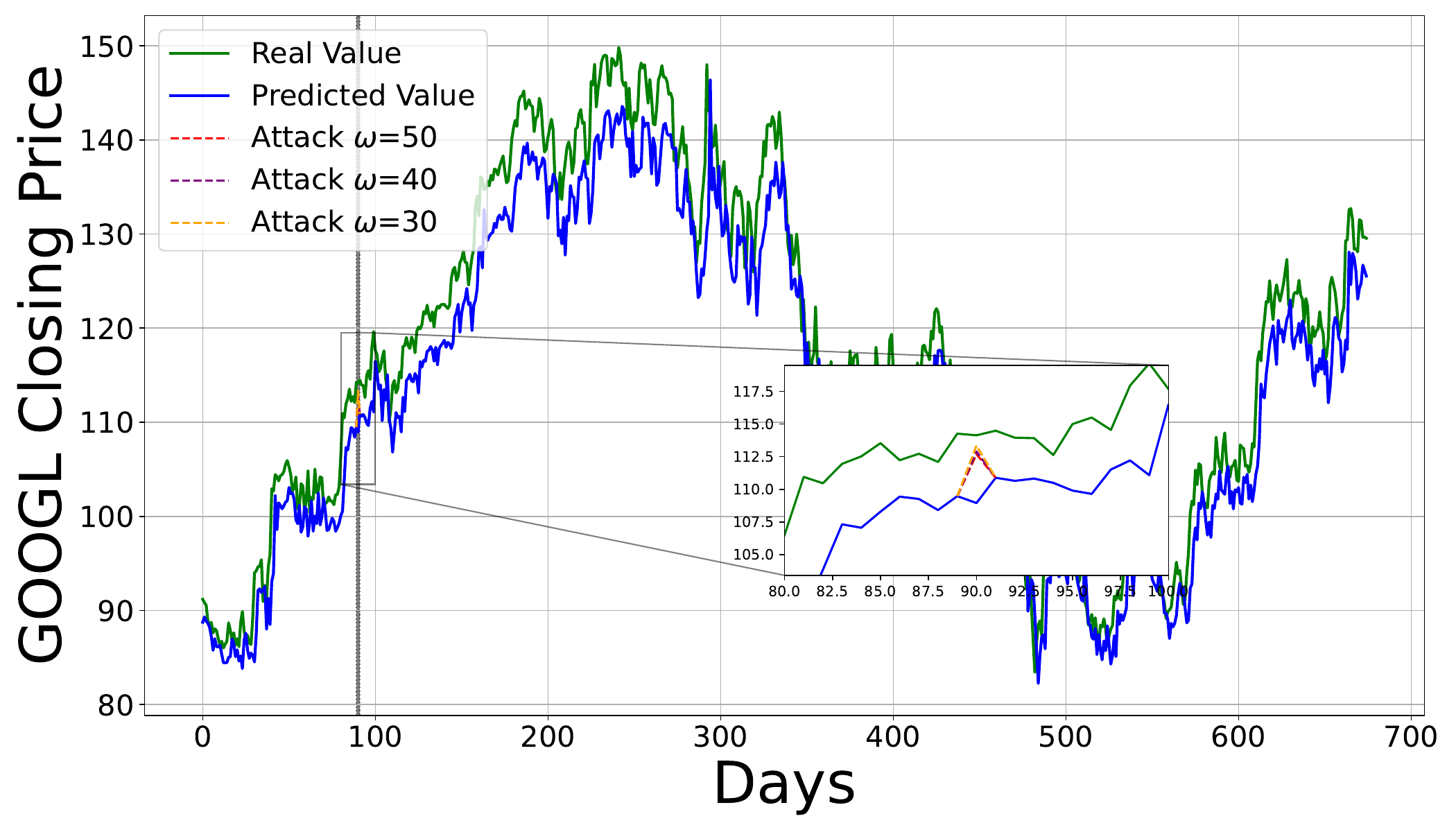}}
        \vspace{-1mm}
        \caption{{\scriptsize \textbf{DL predictor.} \textmd{We introduce a tiny EP (dotted line) on the closing price of GOOGL on day 90. The EP leads the LSTM, when predicting the closing price for day 91, to output a different value. The EP ``disappears'' on day 91.}}}
        \label{sfig:attack_ml}
    \end{subfigure}\hfill
    \begin{subfigure}[!htbp]{0.99\columnwidth}
        \centering
        \frame{\includegraphics[width=0.99\columnwidth]{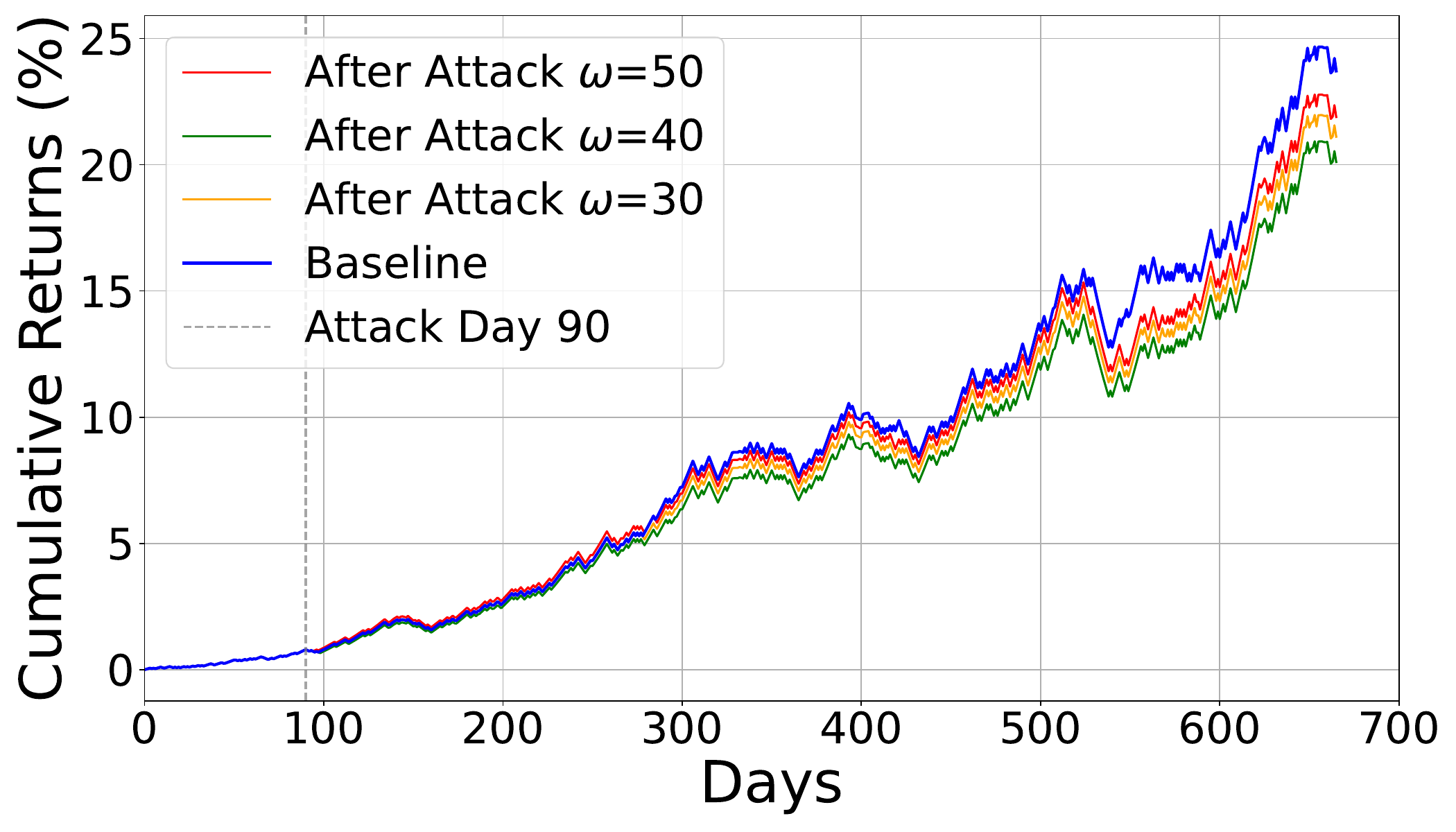}}
        \vspace{-1mm}
        \caption{{\scriptsize \textbf{Whole ATS.} \textmd{Effects of the EP introduced on day 90. Starting from the following day (day 91), the Cumulative Returns of the ATS drops (inducing a monetary loss) w.r.t. the baseline---despite the EP affecting only one day.}}}
        \label{sfig:attack_ats}
    \end{subfigure}
    \vspace{-2mm}
    \caption{\textbf{Exemplary results of an EP.}
    \textmd{\footnotesize We showcase what happens if a DL predictor and overarching ATS are targeted by some of our proposed EP. The blue line represents the baseline performance (y-axis), whereas the others represent the effects of various EPs (targeting the same day, but with different $m$) over our test timeframe (x-axis).}}
    \label{fig:attack}
    \vspace{-4mm}
\end{figure*}

\section{Security Assessment}
\label{sec:assessment}

\noindent
We combine our previous two contributions, and use the ATS developed with our \fmw{} (§\ref{sec:ats}) to assess the impact of perturbations stemming from our threat model (§\ref{sec:threat}). We do so via two case studies, which entail either ``indiscriminate'' (§\ref{ssec:atk_indiscriminate}) or ``targeted'' (§\ref{ssec:atk_targeted}) attacks. We first describe the common setup (§\ref{ssec:atk_setup}).

\subsection{Common Setup}
\label{ssec:atk_setup}

\noindent
Our two case studies share some assumptions. 
First, they both involve the same ATS, i.e., the one we developed in §\ref{ssec:ats_custom} (which we showed is able to generate a profit to its owners---see Fig.~\ref{fig:baseline}). 

Recall that our attacker knows only one stock \smamath{s'\in\mathcal{P}}. For both case studies, \smamath{s'}={\small GOOGL} and, specifically, its closing price: this is a sensible choice, since {\small GOOGL} is a strong asset that is likely to be included in many portfolios, and the closing price is crucial for making educated predictions at the end of the day. In both cases, the attacker also knows that the ATS receives their input data from Yahoo Finance (i.e., the broker): hence, the EP will manipulate the closing price of {\small GOOGL} provided by Yahoo Finance.
Importantly, whenever we apply an EP to the clean data, we ensure that the ATS receives the EP \textit{only} on the attacked day (\smamath{t}). The EP will be \textit{deleted} the following day (because it is overwritten by the new data issued by the broker; see~§\ref{ssec:strategy}): hence, on day \smamath{t+1}, the time-window analyzed by the ATS has \textit{only correct stock values}.

The major difference of our two case studies lies in two crucial aspects of an EP: \textit{when} to launch an EP-based attack (i.e., how to choose \smamath{t}); and \textit{how} to craft the EP (i.e., how to choose \smamath{m}). This will be explained in each case study.

\subsection{First Case Study: Indiscriminate Attack}
\label{ssec:atk_indiscriminate}

\noindent
In this case study, we consider a ``naive'' attacker that does not make advanced consideration on their choice of either \smamath{t} or \smamath{m}. This is useful to examine ``best case'' scenarios for the defender.

\textbf{Setup.} To simulate this scenario in a bias-free way, we craft EP for \textit{all days of the considered test window}. Of course, we will assess each day separately, but the intention is to get a broad understanding of the effects that an EP can have on our ATS on a ``randomly chosen'' day. For the magnitude, we take inspiration from~\cite{gallagher2022investigating}, and consider an EP s.t. \smamath{s'_t}=\smamath{s_t+2\sigma^{\omega}_s}, where \smamath{\sigma^{\omega}_s} is the standard deviation of the value of stock \smamath{s} across the time window \smamath{\omega}. In~\cite{gallagher2022investigating} it is assumed that \smamath{\omega}=\smamath{w}; however, our attacker \textit{does not know} \smamath{w}. Hence, we consider three cases: \smamath{\omega}=(30,40,50), i.e., one case (\smamath{\omega}=50) wherein the attacker is `lucky' and correctly guesses \smamath{w}; and two cases in which the attacker makes a wrong guess. We report in Algorithm~\ref{alg:ephemeral_attack} (at the end of the paper) the pseudocode of our EP-crafting procedure.

\begin{figure*}[!htbp]

\centering
    \begin{subfigure}[!htbp]{0.95\columnwidth}
        \centering
        \includegraphics[width=0.99\columnwidth]{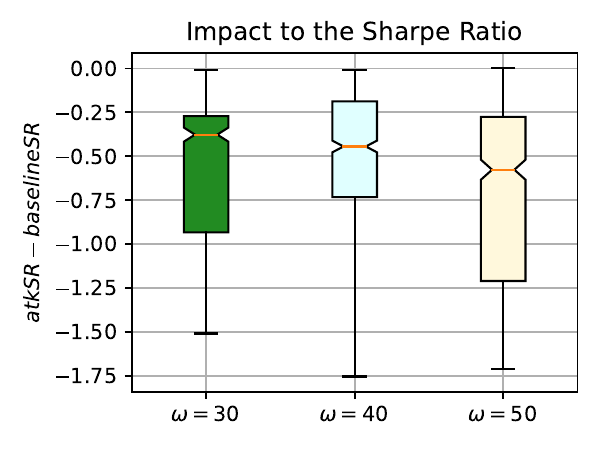}
        \vspace{-3mm}
        \caption{{\scriptsize \textbf{Sharpe Ratio.} \textmd{For 94\% of the cases, the EP lead to a lower SR.}}}

        \label{sfig:sr}
    \end{subfigure}\hfill
    \begin{subfigure}[!htbp]{0.95\columnwidth}
        \centering
        \includegraphics[width=0.99\columnwidth]{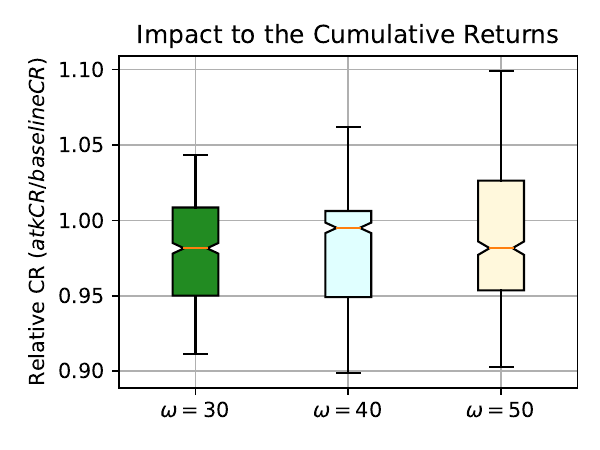}
        \vspace{-3mm}
        \caption{{\scriptsize \textbf{Cumulative Returns.} \textmd{Over 60\% of our EP lead to a lower CR.}}}        
        \label{sfig:cr}
    \end{subfigure}
    \vspace{-2mm}
    \caption{\textbf{Overall impact of our untargeted attacks.}
    \textmd{\footnotesize For each attacked day (of our 666 testing window), we compute: the difference between the Sharpe Ratio (SR) achieved by the ATS at the end of the simulation (i.e., at day 666) with the baseline SR (Fig.~\ref{sfig:sr}); and the ratio between the cumulative returns (CR) achieved by the baseline ATS respect to when it is attacked by an EP. We then plot the distribution of these ``impacts''. For the SR (Fig.~\ref{sfig:sr}) numbers below 0 means that the SR was degraded by the EP; whereas, for the CR (Fig.~\ref{sfig:cr}), numbers below 1 means that the CR was degraded by the EP. Overall, the attack is very successful: for each considered magnitude (\scmath{\omega}=30,40,50) the EP leads to a lower sharpe ratio and inferior cumulative returns in most cases. This is evident by looking at the \textit{notches} of the boxplots (indicating the mean).}}
    \label{fig:boxplot}
    \vspace{-3mm}
\end{figure*}

\textbf{Results.} We craft our ``indiscriminate'' EP for all the 666 days of the test window. We report an exemplary effect of the EP applied to the 90th day of the test window in Figs.~\ref{fig:attack}. Specifically, Fig.~\ref{sfig:attack_ml} shows how the EP affects the LSTM devoted to {\small GOOGL}: we can see that the EP induces this model to make a wrong prediction for a single day (i.e., the 90th); however, after this day, all other predictions are not affected because the EP is sanitized after the ATS receives the ``fixed'' data from the broker. From the perspective of the RMSE, this EP has barely any impact (from 6.3692 to 
6.3662, 6.3662 and 6.3652 for \smamath{\omega}= 50, 40, 30 respectively).
However, the situation changes when we analyse the impact of this EP on the whole ATS, shown in Fig.~\ref{sfig:attack_ats}. We can see that, from day 91, the cumulative returns of the ATS start to deviate (i.e., they are worse) from the baseline. This is due to the ATS making a different trading decision on day 91 (due to the EP): such a decision propagates for the entire length of the test window. While this is the impact for just one day of our test window, we found that similar EP are effective in the wide majority of our considered test window: for \smamath{\omega}= 50, 40, 30 the Sharpe Ratio is lower than the baseline for 628, 646, 635 days out of 666 days, respectively---i.e., over 94\% of our EP decrease the SR compared to the baseline (this is also confirmed by mostly inferior cumulative returns). These results are shown in the boxplots in Fig.~\ref{fig:boxplot}, showing the aggregated impact (relative to the baseline Sharpe Ratio and cumulative returns) of all our EP on our ATS. As it can be seen by the notches of each boxplot (below 0 for the Sharpe Ratio, and below 1 for the Cumulative Returns), the wide majority of our EP substantially decrease the baseline profitability of the ATS.

\subsection{Second Case Study: Targeted Attack}
\label{ssec:atk_targeted}

\noindent
Here, we consider a more sophisticated approach: the attacker \textit{waits} for an opportunity that would allow to craft a ``meaningful'' EP. This is useful to investigate worst case scenarios for the defender.

\textbf{Setup.}
To simulate the above mentioned scenario, we consider an EP launched on days 47, 62, 495, 518, 552, 580 of the test window. On these days important \textit{real-world events} occurred that were related to Google. For instance, on day 62 (i.e., Feb. 8, 2022) the company announced that it would pause hiring for two weeks, whereas on day 518 (i.e., Apr. 17, 2023), its AI chatbot, Bard, gave an incorrect answer in a promotional video. As a matter of fact, the value of {\small GOOGL} dropped significantly on day 63 (by 7.5\%) and day 519 (by 9\%). We explain our reasons for considering these six days in Table~\ref{tab:price_drops}. On such days, a savvy attacker may find it opportune to attack an ATS. We consider two potential scenarios for targeted attacks. First, an attacker trying to ``conceal'' the price drop by reporting the previous value. Second, an attacker (potentially) ``overestimating'' the effects of the news by reporting a very high drop (with respect to the previous day), which we fix at 10\% in our proof-of-concept experiment; of course, different percentages could be possible.

\begin{table}[!htbp]
\centering

\caption{\textbf{Targeted Attacks: chosen days and explanations.}
    \textmd{\footnotesize
    We selected six specific dates (during which the ATS was operational) on which GOOGL stock prices dropped due to real-world events. The table reports the real-world price drop of the closing price of GOOGL (as reported on the following day) with respect to the attacked day.}}
\label{tab:price_drops}
\vspace{-3mm}

\setlength{\tabcolsep}{2pt} 
\renewcommand{\arraystretch}{1}
\resizebox{\columnwidth}{!}{ 
\begin{tabular}{>{\scriptsize}c|>{\scriptsize}l|>{\scriptsize}c|>{\scriptsize}l}
\toprule
\textbf{Day} & \textbf{Date} & \textbf{Drop} & \textbf{Reason} \\ 
\midrule
47 & \scriptsize Jan 24, 2022  & \scriptsize 7.2\%  & \scriptsize The company reported weaker-than-expected earnings.\\
62 & \scriptsize Feb 8, 2022  & \scriptsize 7.5\%  & \scriptsize The company announced that it would be pausing hiring for two weeks.\\
495 & \scriptsize Apr 17, 2023 & \scriptsize 9\%  & \scriptsize Its AI chatbot, Bard, gave an incorrect answer in a promotional video.\\
518 & \scriptsize May 10, 2023  & \scriptsize 4.3\%  & \scriptsize The company reported weaker-than-expected earnings.\\
552 & \scriptsize Jun 13, 2023  & \scriptsize 3.8\%  & \scriptsize The company announced that it would be slowing its pace of hiring.\\
580 & \scriptsize Jul 11, 2023  & \scriptsize 4.1\%  & \scriptsize The company announced that it would be laying off employees.\\
\bottomrule
\end{tabular}
}
\vspace{-3mm}
\end{table}

\textbf{Results.} We measure the impact of the targeted attacks via the relative change (w.r.t. the baseline) in the cumulative returns (CR), shown in Table~\ref{tab:relative_cr_change}. The ``overestimate'' scenario inflicts a substantial loss to the ATS, decreasing the baseline CR by up to 28\%, which makes the ATS almost unprofitable. In contrast, the ``conceal'' scenario is not very malignant (the CR may even increase).

\begin{table}[ht!]
\vspace{-3mm}
    \centering
     \caption{{\footnotesize \textbf{Impact of Targeted EP.}}
    \textmd{\footnotesize CR difference (w.r.t. the baseline) for ``conceal'' and ``overestimate'' atk scenarios.}}
    \vspace{-3mm}
\label{tab:relative_cr_change}
    \small
    \renewcommand{\arraystretch}{0.9}
    \begin{tabular}{lcccccc}
      \toprule
      Atk Day & 47 & 62 & 495 & 518 & 552 & 580 \\
      \midrule
      Conceal & 2.0 & -4.0 & 4.5 & -0.5 & -1.5 & 4.5 \\
      OverEst & -22.0 & -28.0 & -17.5 & -23.5 & -24.5 & -20.0 \\
      \bottomrule
    \end{tabular}
\vspace{-6mm}
\end{table}

\begin{cooltextbox}
\textbf{Takeaway.} We derive two lessons learned from our results.
\textbf{(1)}~\textit{Some EP can lead to an unrecoverable loss by the targeted organization.} Our EP (which, in all cases, are launched in only one day, and involve only a tiny change of the closing price of one stock of the portfolio) lead to a worse SR w.r.t. the baseline.  
\textbf{(2)}~\textit{The effects on the ATS cannot be appreciated with model-only evaluations.} The EP affect the LSTM only for one day. Then, the effects of the EP disappear: the LSTM behaves exactly as if nothing happened---despite the ATS making the organization ``gain less money'' due to the EP. 
\end{cooltextbox} 

\section{Additional Experiments}
\label{sec:additional}

\noindent
We further enrich our security assessment by carrying additional experiments (§\ref{ssec:trading}) and providing a low-level analysis on some ``negative results'' (§\ref{ssec:negative}).

\subsection{More Trading Strategies}
\label{ssec:trading}

\noindent
In our evaluation, we used our EP to attack only one ATS (§\ref{ssec:atk_setup}). Here, we expand our evaluation by considering two additional ATS relying on different trading strategies.

\textbf{Setup.} We consider the same set of 38 LSTM models used to develop our ``primary'' ATS (§\ref{ssec:ats_custom}). However, instead of using the output of these models as input to an ATS that leverages the ``moving average crossover'' trading strategy, we consider two different trading strategies---both adopted also by prior work (e.g.,~\cite{kumar2022comparative, zheng2022research,yan2023enhanced}): 
\begin{itemize}[leftmargin=*]
    \item \textit{Rate of Change~\cite{roc}:} this computes a reference metric called ``Rate of Change'' for every stock in the portfolio. By following best practices~\cite{roc}, we compute such a metric as follows: given a stock \smamath{s}, its ``Rate of Change'' (\smamath{ROC_{s}^{t}}) on day \smamath{t} is \smamath{ROC_{s}^{t}=\frac{s_t-s_{t-14}}{s_{t-14}}*100}, with \smamath{s_t} being the value of stock \smamath{s} at time \smamath{t}. To trigger a buy signal, the \smamath{ROC} must be above 1; for a sell signal, the \smamath{ROC} should be below -1; otherwise, no decision is made (i.e., default to hold).
    
    \item \textit{Bollinger Bands~\cite{bollinger}:} the fundamental principle of this trading strategy is to make a decision depending on where the predicted value of a stock falls in a given reference ``band''. In our case, we use the default parameters~\cite{kothapalli2023predicting}: we compute the 20-day moving average of the closing price of each stock, and choose an upper/lower bound by adding/subtracting twice its standard deviation: if the prediction of the model is above the ``upper'' bound, then this triggers a buy signal; if it is below the ``lower'' bound, then this triggers a sell signal; otherwise, it is a hold.
\end{itemize}
We set the same initial conditions as in our ``primary'' ATS for both of these additional ATSs, and we let them run over the same test period. At the end of the simulation, in the absence of attacks, the Sharpe Ratio of both of these ATS is above 0. Specifically, for the ATS using Bollinger Bands, its Sharpe Ratio=0.26, whereas for the ATS using the Rate of Change, its Sharpe Ratio=0.72. In both cases, however, the Sharpe Ratio is lower than the one achieved by our ``primary'' ATS (see Fig.~\ref{sfig:sharpe_ratio}). In contrast, the cumulative returns of the Bollinger Bands ATS are almost negligible (-0.3\%) despite its positive Sharpe Ratio (this is because this trading strategy adopts a very conservative approach); whereas for the Rate of Change they are higher (49\%) than our ``primary'' ATS. This is because the Rate of Change is a very risky trading strategy which, in our case, ``paid off''; however, such a huge risk factor leads to a lower Sharpe Ratio.

\begin{figure}[!htbp]

\centering
    \begin{subfigure}[!htbp]{0.48\columnwidth}
        \centering
        \includegraphics[width=0.99\columnwidth]{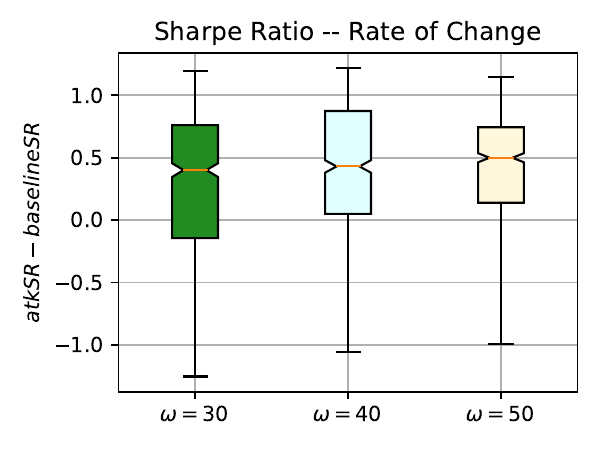}
        \vspace{-2mm}
        \caption{{\scriptsize \textbf{Sharpe Ratio.} \textmd{For 19.5\% of the cases, the EP lead to a lower SR.}}}

        \label{sfig:sr_2}
    \end{subfigure}\hfill
    \begin{subfigure}[!htbp]{0.48\columnwidth}
        \centering
        \includegraphics[width=0.99\columnwidth]{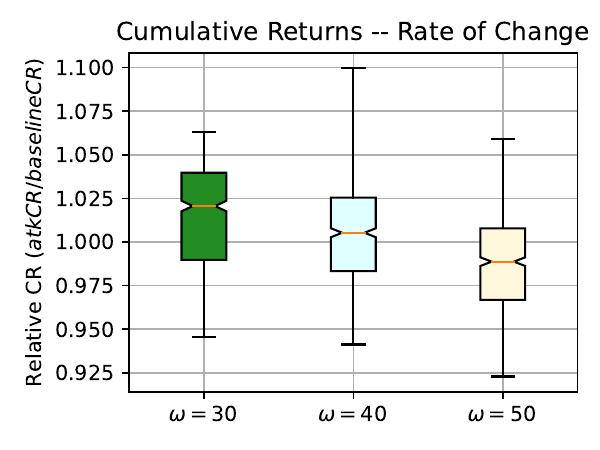}
        \vspace{-2mm}
        \caption{{\scriptsize \textbf{Cumulative Returns.} \textmd{Over 49\% of our EP lead to a drop in the baseline CR.}}}        
        \label{sfig:cr_2}
    \end{subfigure}
    \vspace{-2mm}
    \caption{\textbf{Overall impact of EP against the ATS using the Rate of Change trading strategy.}
    \textmd{\footnotesize For each attacked day (of our 666 testing window), we compute: the difference between the Sharpe Ratio (SR) achieved by the ATS at the end of the simulation (i.e., at day 666) with the baseline SR; and the ratio between the cumulative returns (CR) achieved by the baseline ATS respect to when it is attacked by an EP. We then plot the distribution of these ``impacts''.}}
    \label{fig:boxplot_2}

    \vspace{-3mm}
\end{figure}

\begin{figure}[!htbp]
\vspace{-3mm}
\centering
    \begin{subfigure}[!htbp]{0.48\columnwidth}
        \centering
        \includegraphics[width=0.99\columnwidth]{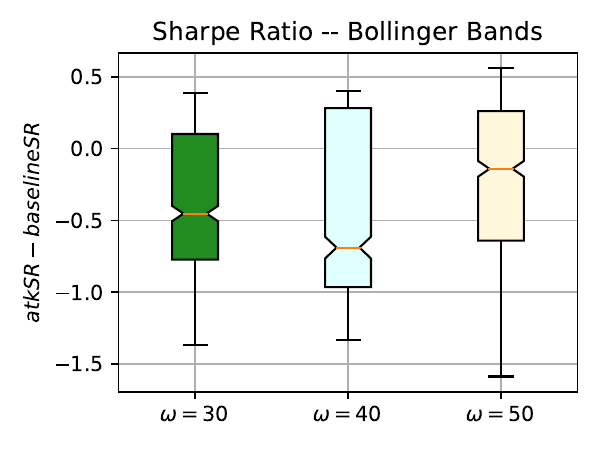}
        \vspace{-2mm}
        \caption{{\scriptsize \textbf{Sharpe Ratio.} \textmd{For 62\% of the cases, the EP lead to a lower SR.}}}

        \label{sfig:sr_3}
    \end{subfigure}\hfill
    \begin{subfigure}[!htbp]{0.48\columnwidth}
        \centering
        \includegraphics[width=0.99\columnwidth]{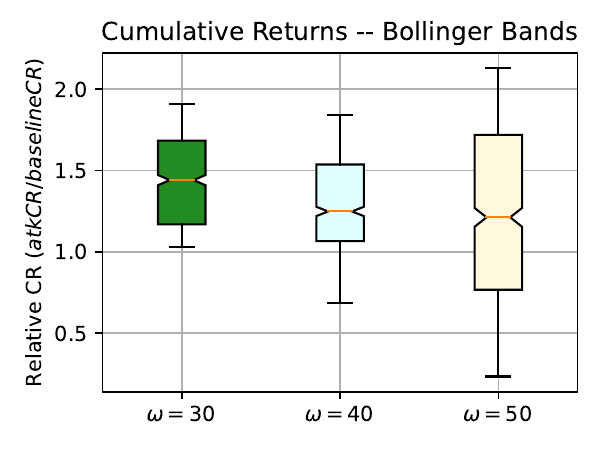}
        \vspace{-2mm}
        \caption{{\scriptsize \textbf{Cumulative Returns.} \textmd{For 18.8\% of the cases, the EP lead to a lower CR.}}}        
        \label{sfig:cr_3}
    \end{subfigure}
    \vspace{-2mm}
    \caption{\textbf{Overall impact of EP against the ATS using the Bollinger Bands trading strategy.}
    \textmd{\footnotesize For each attacked day (of our 666 testing window), we compute: the difference between the Sharpe Ratio (SR) achieved by the ATS at the end of the simulation (i.e., at day 666) with the baseline SR; and the ratio between the cumulative returns (CR) achieved by the baseline ATS respect to when it is attacked by an EP. For Fig.~\ref{sfig:cr_3}, we normalized the values between 0 and 1 before making the plot (since the baseline CR was a negative number)}}
    \label{fig:boxplot_3}
    
\end{figure}

\textbf{Results.} Our EP are agnostic of the ATS (and of its underlying DL models), hence we test these two ATS against the exact same EP used against our ``primary'' ATS. The complete results are provided in Fig.~\ref{fig:boxplot_2} (for the Rate of Change trading strategy) and in Fig.~\ref{fig:boxplot_3} (for the Bollinger Bands trading strategy). To allow a high-level comparison, both of these figures have the same structure of Fig.~\ref{fig:boxplot}. We can see some intriguing results.
\begin{itemize}[leftmargin=*]
    \item \textit{EP vs Rate of Change.} In terms of cumulative returns, over 49\% of our EP led to a net loss with respect to the baseline (see Fig.~\ref{sfig:cr_2}). In contrast, only 20\% of our EP led to a reduction of the Sharpe Ratio (see Fig.~\ref{sfig:sr_2}), with an absolute (negative) difference that could go below -1.2 than the baseline. Such a discrepancy is due to the high-risk nature of this strategy: despite inducing a lower net revenue, the Sharpe Ratio was not excessively affected.
    \item \textit{EP vs Bollinger Bands.} For the cumulative returns, 18.8\% of our EP led to a net loss with respect to the baseline (see Fig.~\ref{sfig:cr_3}). Intriguingly, however, 62\% of our EP led to a drop in the Sharpe Ratio. And, in fact, the notches of the boxplots in Fig.~\ref{sfig:sr_3} are all below the saddle point (of 0). These effects can be explained by the fact that the Cumulative Returns of this trading strategy were almost 0 without any EP, hence it makes sense that the introduction of an EP may not have excessive effects on this metric (and, actually, induce the ATS to gain more money, as remarked by the notches being over 1). While this ATS may appear to be ``more robust'' against EP than the others, its baseline profitability was also vastly inferior.
\end{itemize}
We can conclude that evaluating such system-wide properties of EP is fundamental to gauge their practical effects. Importantly: the two ATS considered in this expanded assessment were relying on the \textit{exact same DL models} of our ``primary'' ATS. This shows that, e.g., in some cases the perturbations can have little impact (e.g., against the ATS using the Rate of Change). 

\subsection{Negative Result: A lesson learned}
\label{ssec:negative}

\noindent
Here, we use one of the EP of our ``primary'' ATS to further highlight the importance of assessing the system-wide properties of adversarial perturbations (including, of course, our EP) against ATS.

\begin{figure}[!htbp]
    \vspace{-1mm}
    \centering
    \includegraphics[width=0.95\columnwidth]{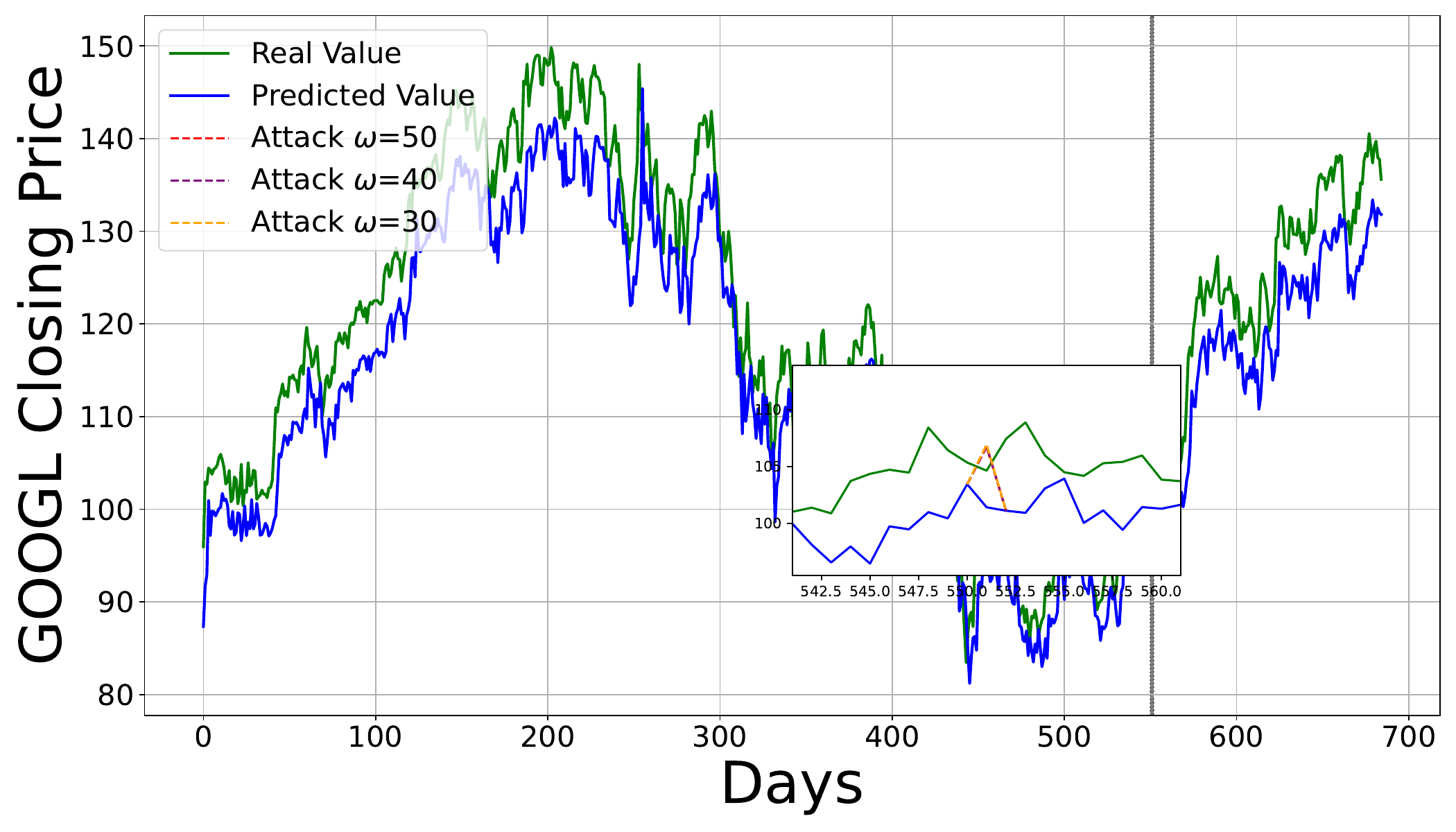}
    \vspace{-4mm}
        \caption{\textbf{Negative result.}
    \textmd{\footnotesize This EP, introduced for day 551, affected the LSTM model (as shown) but did not lead to any change in the profitability of our ``primary'' ATS.}}
    \label{fig:negative}
    \vspace{-3mm}
\end{figure}

Specifically, we focus on the EP that we craft for day 551 of our simulation. We showcase the effects of such an EP on the LSTM (analysing {\small GOOGL}) in Fig.~\ref{fig:negative}. We can see that the predicted value of the LSTM for day 552 is around 101, but introducing the EP on day 551 leads the model to predict a value that is about 106. The value of the stock on day 551 was 105, i.e., \textit{below} the value predicted due to the EP, but \textit{above} the value predicted without an EP. Yet, despite the perturbation clearly leading to a wrong prediction by the considered model, such an EP \textit{had no impact} on the overarching system---neither in terms of the Sharpe Ratio (which was 1.37, i.e., the same as the baseline), nor for the cumulative returns (which were 24\%, i.e., they slighly increased from the baseline 23.7\%).

\noindent
We use such a negative result as a scaffold for two considerations.
\begin{itemize}[leftmargin=*]
    \item First, the EP led the LSTM to produce a different output---hence, from an ``adversarial ML point of view'', the EP was successful. However, such an EP had no impact on the overarching ATS. 
    
    \item Second, the EP induced a similar effect as those of adversarial perturbations envisioned by prior work. For instance, in~\cite{nehemya2021taking,dang2020adversarial}, the goal was to craft a perturbation that induced the model to output a wrong prediction in terms of buy/sell, with the logic that ``if the model predicts that the price drops, then it is a sell signal---hence, my perturbation should induce the model to predict that the price increases'' (and viceversa). This is a sensible goal: as a matter of fact, the trading strategies of our ATS also leverage similar rationales. However, as we showed, such an EP has no impact on our ATS---and prior works did not evaluate the impact of their perturbations on the overarching system.
\end{itemize}
The reason why this EP led to no impact is because, on that day, the ATS decided that it was better not to do anything with the {\small GOOGL} stock---defaulting to ``hold'' with or without the EP. For instance, the baseline system (which predicted a price drop) may not have had any {\small GOOGL} stock to sell; whereas the ``attacked'' system (seeing a price increase) may not have had enough resources to purchase a {\small GOOGL} stock. Moreover, even when a buy or sell signal is triggered, this does not mean that the ATS will carry out such an operation: besides available resources, there is also the risk factor to consider. Perhaps, in the case of buy, the ATS may have preferred to buy another stock for which the corresponding LSTM (which was not attacked) predicted a huge increase.

\vspace{1mm}
{\setstretch{0.8}
\textbox{{ \textbf{Lesson Learned.} ATS are complex systems and their profitability depends on a plethora of factors. Carrying out model-only evaluations may overestimate the efficacy of a given attack. This is why we advocate future work to expand their scope and carry out system-wide assessments---which are enabled by our \fmw{}.}}
}

\section{Discussion}
\label{sec:discussion}

\noindent
Our paper has addressed security issues that may affect the ATS deployed by real organizations, and seeks to provide a foundation for further research in this field. Here, we reflect on our contributions by outlining the opinion of experts in the financial industry~(§\ref{ssec:expert}), by discussing the limitations of our work~(§\ref{ssec:limitations}), and by providing additional analyses of our threat model (§\ref{ssec:analysis}).

\subsection{User Study with Experts}
\label{ssec:expert}

\noindent
In Sept--Dec 2023, we reached out to 7 experts who work in the FinTech industry and have experience in the field of AI for finance, including ATS. We presented our research to the experts and asked for their opinion via three questions:
\begin{itemize}[leftmargin=*,noitemsep,topsep=0pt]
    \item Is our ATS practical in terms of its performance?
    \item Does our threat model reflect a feasible scenario?
    \item Does our EP represent a security risk?
\end{itemize}
We acknowledge that, due to the small size of our sample we cannot claim that the answers we received fully reflect the opinion of a broad range practitioners worldwide. For confidentiality reasons, we cannot provide the complete answers we received or further details about our respondents.

Overall, the responses of our interviewees were positive for all three questions. The answers confirmed that the positive Sharpe Ratio would justify the deployment of such an ATS for real-life trading, and that our chosen portfolio is reasonable. It was also noted that our threat model constitutes a tangible risk; some experts even acknowledged the likelihood of man-in-the-middle attacks between brokers and ATS and hence the risk of EP. Finally, some experts even remarked that organizations may use ATS in a ``fire and forget'' fashion, i.e., once the ATS is shown to yield some profit in the short-term, it will be deployed for several months until its performance is re-assessed. This implies that a hard-to-identify EP would adversely affect the trading decisions for a long time.

\subsection{Limitations and Disclaimers}
\label{ssec:limitations}

\noindent
Our \fmw{} framework (§\ref{sec:ats}) seeks to establish the means for security assessments of DL-based ATS. To enable a fine-grained control over its functionality and for better understanding of the operational characteristics of ATS, we have chosen to implement such a framework from scratch despite the existence of toolkits/platforms for ATS simulation, e.g., Backtrader~\cite{backtrader} or Quantiacs~\cite{quantiacs}. We acknowledge that off-the-shelf tools may attain better real-world fidelity or offer more advanced capabilities for portfolio management. A comprehensice comparison of our framework with such tools is beyond the scope of this work. 

Our security assessment is meant to provide a proof-of-concept evaluation of the potential impact of our threat model~(§\ref{sec:assessment}). We cannot claim that any \emph{operational ATS} is affected by EP in the same way as shown in our experiments; nor we claim that the attack is ``universally devastating''. However, research on the security of AI in finance is scarce (as we showed in §\ref{sec:related}), and our findings reveal that even when the impact on the model is negligible, the overarching system may still be significantly affected. Such a factual consideration was not apparent from prior work (e.g.,~\cite{nehemya2021taking}).

Finally, our focus is in DL-based ATS. However, there are other ways to develop predictive models for stock price prediction (e.g., ARIMA~\cite{siami2018comparison,pirani2022comparative}): security assessment of these methods is outside our scope---but our resources enable one to evaluate these techniques, too (we provide examples in our repository~\cite{ourRepo}).

\subsection{Additional Analyses on our Threat Model}
\label{ssec:analysis}

\noindent
We analyse some aspects of our threat model that can be used to extend our findings, but also as intriguing avenues for future work.

\vspace{1mm}
\noindent
\textbf{Robustness considerations.}
Our EP entails applying an imperceptible change to a single data point. Albeit we assume that such a change is introduced with a malicious purpose, it is entirely possible that such perturbations are caused by ``neutral'' events. For instance, a broker's may have received  wrong data which is transmitted to the ATS. Hence, evaluating the effects of EP can be useful not only as a security assessment, but also as a means to investigate the overall robustness of ATS against data perturbations. We leave development of countermeasures against EP to future work (facilitated by our tools). Given our results, we believe a pragmatic defense to EP to be an intriguing avenue for this research domain. 

\vspace{1mm}
\noindent
\textbf{More knowledge of the Portfolio}
Our threat model assumes the attacker knows only one stock in the portfolio (\smamath{s'\in\mathcal{P}}), but this can be extended. The attacker could instead know a subset \smamath{S'\subseteq\mathcal{P}}, allowing them to choose multiple stocks \smamath{s \in S'} for crafting EPs. This increased knowledge could lead to stronger effects. However, obtaining such information is costly (i.e., the attacker cannot access the ATS from within). The attacker might infer portfolio details through privacy violations or insider actions~\cite{woods2021sok_fc}, which are avenues for future work.

\vspace{1mm}
\noindent
\textbf{Alternative Goals and Strategies}
Our threat model envisions a constrained attacker aiming to harm an organization by reducing the yield of its ATS.  Other methods exist to achieve this goal. For instance, an attacker with some control over communications between the broker and the ATS might launch a Denial of Service attack, preventing trades on a specific day. However, such a strategy is conspicuous, and the organization would likely react---which may lead to the attacker being detected. 
Conversely, it is also possible that the attacker uses the EP for ``poisoning'' a DL model: if an EP is not sanitized, it can be stored by the ATS. Such an EP can lead to changes in the predictions of multiple days (i.e., the EP can potentially affect all the following \smamath{w} days), but it can also have more collateral effects if the EP is used in the re-training process of the models. Our \fmw{} enables future work to investigate these ancillary adversarial scenarios.

\section{Conclusions}
\label{sec:conclusions}

\noindent
We elucidated the vulnerability of DL systems in finance to a novel, subtle type of adversarial perturbation, which we defined as ``ephemeral perturbation'' (EP). Compared to the limited prior works that studied the security of DL applications in computational finance, our research showed that our EP (which stem from a more constrained attacker than those envisioned in prior threat models) have an almost negligible impact on the performance of the affected DL model, but can reduce the profitability of a DL-powered algorithmic trading system (ATS). The latter finding is of crucial importance: security assessments studying the impact of data perturbations on the whole system are scarce in the financial context.

Future work should pay more attention to this application domain of DL---e.g., by devising countermeasures against our proposed EP; or assessing the effects of EP in high-frequency contexts, whose transactions occur every second (instead of daily); or even by relaxing some of the underlying assumptions of our proposed threat model, and studying their effects on the ATS end-to-end. Our proposed security assessment framework (\fmw{}) and our resources enable downstream research to carry out all such analyses. 

\vspace{1mm}
{\setstretch{0.8}
\textbox{\textbf{Open source.} Our experiments are reproducible, and our resources are available in a publicly available repository~\cite{ourRepo}, in which we also report additional experimental results and visualizations of our assessment, as well as ``how-to'' guides.}}

\section*{Acknowledgement}
\noindent
The authors want to thank the organizers and attendees of WACCO'24 for the constructive feedback, and the Hilti Foundation for funding this research (which has also been funded by the SAMLAF project).

\begin{algorithm}

\caption{Applying an EP to time-series data of a given stock}
\label{alg:ephemeral_attack}
\begin{algorithmic}[1]
\Require Time-series data {\small $\mathcal{TS}$}, EP magnitude $m$, time-series window size $w$, window-size expected by the attacker $\omega$
\Ensure Set of adversarially perturbed time-series data {\small $\overline{\mathcal{TS}'}$}
\vspace{2mm}
\State \textit{//~Procedure for generating an adversarial time-series}
\Function{ApplyEP}{\smacal{TS}, $t, m, \omega$}
    \State {\small $\mathcal{TS}'$} $\gets$ \smacal{TS} \Comment\textit{{\textcolor{blue}{Copy original time-series}}}
    \If{$m = \text{std}$} \Comment\textit{\textcolor{blue}{Our Indiscriminate Attack (§\ref{ssec:atk_indiscriminate})}}
        \State $\sigma \gets \text{std}(\mathcal{TS}[t - \omega:t])$ \Comment\textit{\textcolor{blue}{{std.dev.-based EP}}}
        \State {\small $\mathcal{TS}'[t]$} $\gets$ {\small $\mathcal{TS}[t]$} + $2 \cdot \sigma$
    \Else \Comment\textit{\textcolor{blue}{Our Targeted Attack (§\ref{ssec:atk_targeted})}}
        \State {\small $\mathcal{TS}'[t]$} $\gets$ custom \Comment\textit{{\textcolor{blue}{Ad-hoc crafting of EP}}}
    \EndIf

    \State \Return {\small $\mathcal{TS}'$}
\EndFunction

\vspace{0.2cm}

\State {\small $\overline{\mathcal{TS}'}$} $\gets \emptyset$
\For{($t=w$, $t\in$~\smacal{TS}, $t=t+1$)}
    \State {\small $\overline{\mathcal{TS}'}$} $\gets$ {\small $\overline{\mathcal{TS}'}$} $\cup$ \textsc{ApplyEP}(\smacal{TS}, $t, m, \omega$)
\EndFor
\Statex \textcolor{brown}{(\textbf{Note:} the magnitude $m$ can be freely defined by an user)}

\end{algorithmic}

\end{algorithm}

\bibliographystyle{ACM-Reference-Format}



\end{document}